\documentclass[aps,prb,twocolumn,superscriptaddress]{revtex4-1}

\usepackage{graphicx}        
\usepackage{amsmath, amssymb}
\usepackage{bm}
\usepackage{siunitx}
\usepackage{braket}
\usepackage[colorlinks=true, linkcolor=blue, citecolor=blue]{hyperref}
\usepackage[utf8]{inputenc}
\renewcommand{\vec}[1]{\mathbf{#1}}
\graphicspath{{./}}

\begin{document}

\title{Topology driven g-factor tuning in type-II quantum dots}


\author{J. M. Llorens}
\affiliation{Instituto de Micro y Nanotecnología, IMN-CNM, CSIC (CEI UAM+CSIC) Isaac Newton, 8, E-28760, Tres Cantos, Madrid, Spain}

\author{V. Lopes-Oliveira}
\affiliation{Universidade Estadual de Mato Grosso do Sul, Dourados, Mato Grosso do Sul 79804-970, Brazil}

\author{V. López-Richard}
\affiliation{Departamento de Física, Universidade Federal de São Carlos, São Carlos, São Paulo 13565-905, Brazil}

\author{E. R. Cardozo de Oliveira}
\affiliation{Departamento de Física, Universidade Federal de São Carlos, São Carlos, São Paulo 13565-905, Brazil}

\author{L. Wewiór}
\affiliation{Instituto de Micro y Nanotecnología, IMN-CNM, CSIC (CEI UAM+CSIC) Isaac Newton, 8, E-28760, Tres Cantos, Madrid, Spain}

\author{J. M. Ulloa}
\affiliation{Institute for Systems based on Optoelectronics and Microtechnology (ISOM), Universidad Politécnica de Madrid, Ciudad Universitaria s/n, 28040 Madrid, Spain}

\author{M. D. Teodoro}
\affiliation{Departamento de Física, Universidade Federal de São Carlos, São Carlos, São Paulo 13565-905, Brazil}

\author{G. E. Marques}
\affiliation{Departamento de Física, Universidade Federal de São Carlos, São Carlos, São Paulo 13565-905, Brazil}

\author{A. García-Cristóbal}
\affiliation{Instituto de Ciencia de Materiales (ICMUV), Universidad de Valencia, Paterna E-46980, Spain}

\author{G.-Q. Hai}
\affiliation{Instituto de Física de São Carlos, Universidade de São Paulo, São Carlos, São Paulo 13560-970, Brazil}

\author{B. Alén}
\affiliation{Instituto de Micro y Nanotecnología, IMN-CNM, CSIC (CEI UAM+CSIC) Isaac Newton, 8, E-28760, Tres Cantos, Madrid, Spain}
\email[E-mail address:]{benito.alen@csic.es}

\date{\today}

\begin{abstract}
We investigate how the voltage control of the exciton lateral dipole moment induces a transition from singly to doubly connected topology in type-II InAs/GaAsSb quantum dots. The latter causes visible Aharonov-Bohm oscillations and a change of the exciton $g$-factor which are modulated by the applied bias. The results are explained in the frame of realistic $\vec{k}\cdot\vec{p}$ and effective Hamiltonian models and could open a venue for new spin quantum memories beyond the InAs/GaAs realm.
\end{abstract}


\maketitle

\section{Introduction}
III-V semiconductor quantum dots (QDs) are a fundamental resource for quantum optical information technologies, from integrated quantum light sources to quantum optical processors.\cite{michler_quantum_2017} Within this broad field, most knowledge arises from the InAs/GaAs system, with many other compounds being still relatively unexplored. III-Sb QDs and rings are a good example. They can be grown by several epitaxial methods and can strongly emit in all the relevant telecom bands, yet, III-Sb infrared quantum light sources are to be developed. III-Sb compounds also have the largest $g$-factor and spin-orbit coupling (SOC) constant of all semiconductors but, with the exception of InSb nanowires,~\cite{nadj-perge_spectroscopy_2012} this advantage has yet to be exploited in spin based quantum information technologies. One advantage of III-V-based quantum nanostructures is that optical initialization and read-out can be done in a few nanoseconds using, for instance, the singly charged exciton lambda system with electrons~\cite{atature_quantum-dot_2006} or holes~\cite{brunner_coherent_2009}. Using the hole brings the additional benefit of its p-orbital character, meaning that the overlap between the atomic orbitals and the nucleus is significantly smaller than for the s-like electrons, diminishing the decoherence caused by the hyperfine interaction.~\cite{brunner_coherent_2009,vidal_hyperfine_2016}

For quantum information processing in the solid state, the possibility to modify the $g$-factor through an external bias is its most outstanding resource.~\cite{nowack_coherent_2007} Thanks to spin orbit coupling terms in the Hamiltonian, the application
of an electric field can modify the $g$-factor magnitude and sign through changes in the orbital part of the QD wavefunctions~\cite{pingenot_method_2008,andlauer_electrically_2009,
pingenot_electric-field_2011}. A similar effect can be obtained changing the elastic strain around the QD.~\cite{tholen_strain-induced_2016}
Since the $g$-factor and wavefunctions are anisotropic, the observed modulation is different in Faraday~\cite{klotz_observation_2010,jovanov_observation_2011,ares_nature_2013,corfdir_tuning_2014} and Voigt~\cite{godden_fast_2012} configurations. With InAs/GaAs QDs, external bias in the range of tens of kV cm$^{-1}$ is necessary and, to prevent electron tunneling out of the QD, the introduction of blocking barriers is advisable.~\cite{bennett_voltage_2013,prechtel_electrically_2015} Alternately, one could use semiconductor nanostructures with larger spin orbit coupling constants to produce larger modulations at lower bias.

\begin{figure}[h]
        \centering
        \includegraphics[width=\columnwidth]{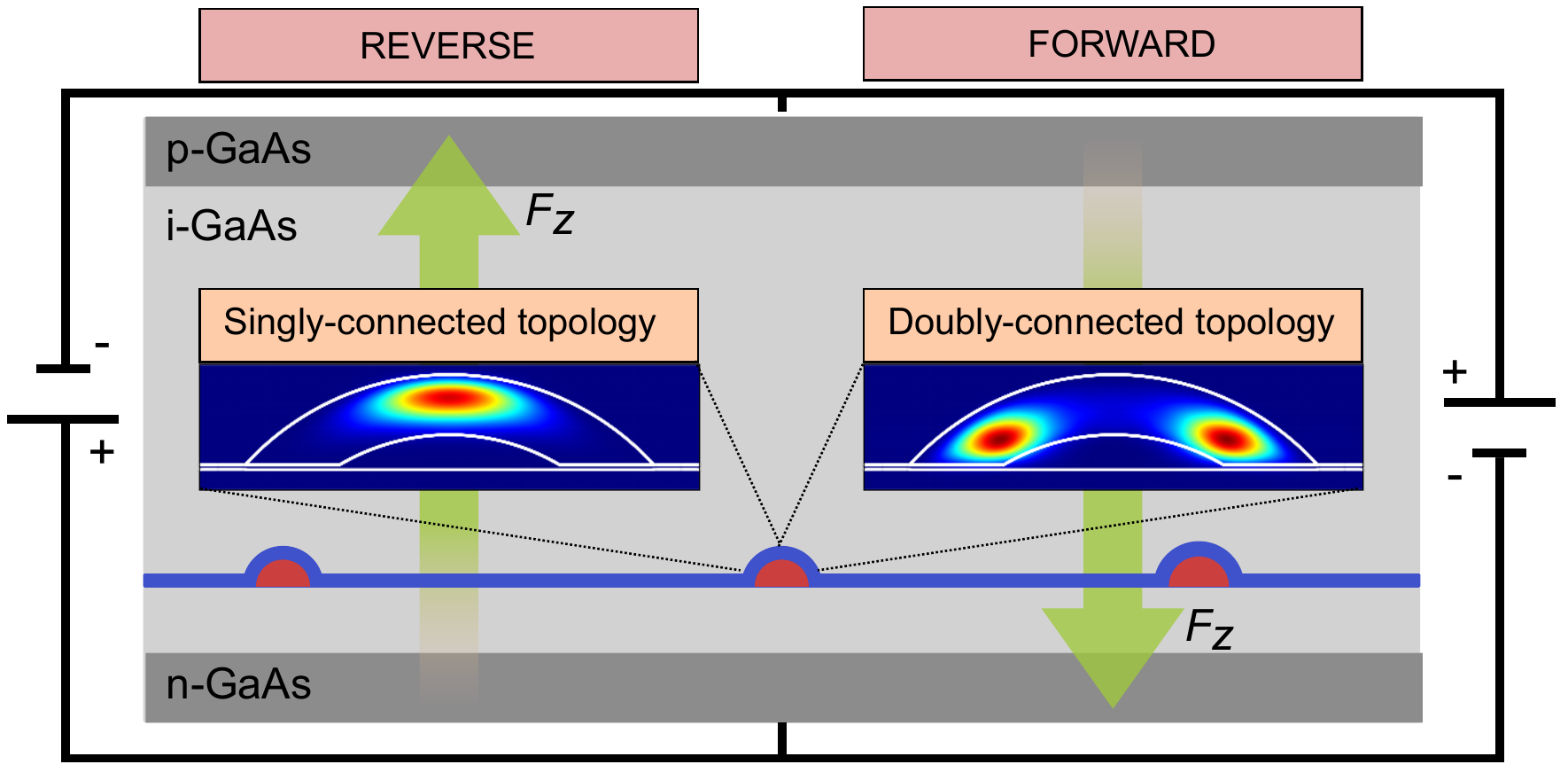}
        \caption{Depiction of the p-i-n diode operational principle acting on
        the hole wave function. Depending on the polarization of bias the applied
electric field ($F_z$) points upwards (left) or downwards (right) changing
the topology of the wave function from singly-connected to doubly-connected.
The size of the elements is not to scale.}
        \label{fig:Intro}
\end{figure}

Meanwhile, quantum nanostructures with a ring-shaped topology for electrons, holes or both and type-I or type-II confinement have been the object of intense theoretical and experimental research.~\cite{Fomin_QR_Book_1st_eddition,
lorke_spectroscopy_2000,fuhrer_energy_2001,ribeiro_aharonov-bohm_2004,dias_da_silva_polarization_2005,
kleemans_oscillatory_2007,Kuskovsky2007,sellers_aharonov-bohm_2008,Marcio_AB_2010,miyamoto_excitonic_2010,kim_observation_2016}.
When two particles travel around these quantum rings (QR) circumventing a magnetic
flux, the Aharonov-Bohm (AB) effect changes the relative phase factor of their
wave functions.~\cite{aharonov_significance_1959} The optical Aharonov-Bohm
effect (OABE) can be detected at optical frequencies through the relative
changes in the electron and hole orbital angular momenta. 
At zero magnetic field, electron-hole exchange interaction typically leads to a dark ground state doublet split by few hundred $\mu$eV from the bright exciton doublet. With increasing applied magnetic field, higher energy states with non-zero orbital angular momentum cross this quadruplet becoming the new ground state of the system. These states are hence optically forbidden and must produce a fade-out of the emitted intensity.~\cite{Govorov_AB_2002} In actual samples such a reduction on the emitted light is not observed because of the reduced symmetry of the system, i.e. the orbital angular momentum is not a good quantum number any longer. As a consequence, oscillations in energy and intensity are observed instead of a quench.~\cite{Govorov_AB_2002} The evolution of charged excitons becomes more complex in either case.~\cite{bayer_optical_2003,okuyama_optical_2011,llorens_wave-function_2018}

Given the different orbital confinement found in singly and doubly connected potentials, nanostructures with electrically tunable topology might be of great interest to couple spin and orbital degrees of freedom.~\cite{Llorens2015,llorens_wave-function_2018} In the following, we investigate such possibility focusing on type-II InAs/GaAsSb QDs grown on GaAs
and embedded in a p-i-n diode structure. For Sb molar fractions beyond 16\si{\percent}, the band alignment changes from type-I to type-II, with the electron confined
inside the InAs and the hole delocalized in the GaAsSb overlayer.~\cite{akahane_long-wavelength_2004,ripalda_room_2005,
liu_long-wavelength_2005}. The resulting wave function configuration brings two
important assets. Firstly, the weak localization of the hole results in a
larger exciton polarizability or, in other words, a higher tunability of the
exciton energy. Voltage control of the exciton dipole moment in the vertical
direction thus allows large tuning of the radiative
lifetime.~\cite{Llorens2015,heyn_field-controlled_2018} Secondly, we shall see that, pushing or pulling
the hole against the bounding interfaces of the overlayer, also brings a
topological change in the hole ground state wave function 
as schematically shown in Figure~\ref{fig:Intro}. By this manipulation we are in fact engineering the wave function density of probability from dot- to ring-like as originally discussed in Ref.~\onlinecite{kleemans_oscillatory_2007}

The rest of the paper is organized as follows. The details of the sample and optical characterization are summarized in Section~\ref{sec:Exp}. We present in Section~\ref{sec:VBz} the magneto-optical photoluminescence results. We first find that both, the diamagnetic coefficient and the $g$-factor, can be modulated by the applied bias. Then, we report oscillations of the emission intensity and the degree of
circular polarization ($DCP$) with the applied magnetic field which occur only under forward bias.  We discuss these effects in
Section~\ref{sec:Discusion}, relying on two different theoretical models. The
first one is described in Section~\ref{subsec:kpEzBz}. It consists of
a multiband $\vec{k}\cdot\vec{p}$ axisymmetric model which
provides a quantitative evolution of the orbital related effects:
the diamagnetic coefficient dependence and the intensity oscillation. 
The limitation of the axial symmetry of this
model is complemented with an effective mass model in
Section~\ref{subsec:asymmetry}. It is based on a parabolic confinement which can
analytically evolve from a dot-like to a ring-like potential and includes an
eccentricity parameter. From the solutions of this model, we discuss the spin
related results ($g$-factor and $DCP$) in Section~\ref{sec:SOC} by including
the Rashba contribution in the Hamiltonian. 

%
\begin{figure*}[tb]
\includegraphics[scale=.85]{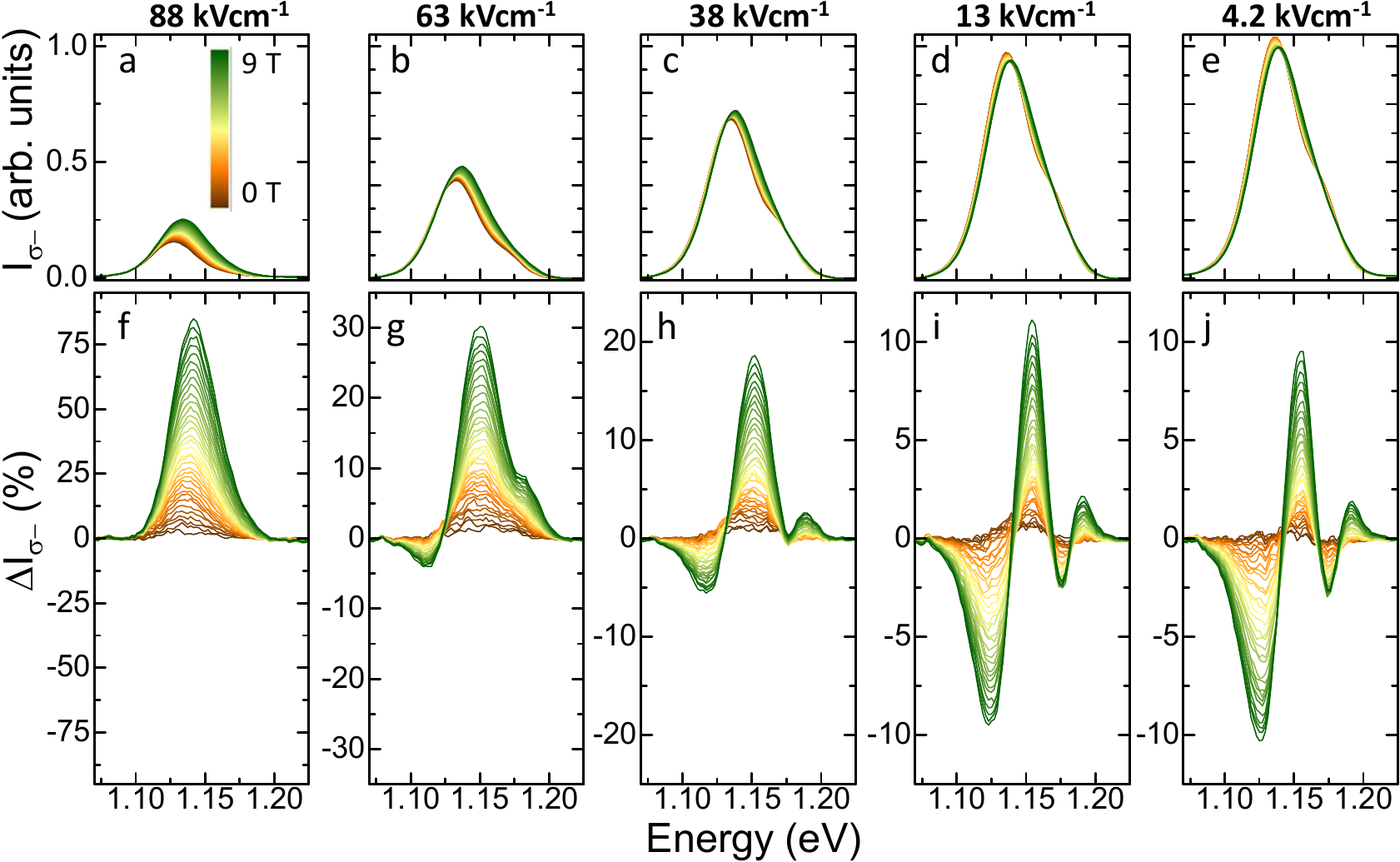}
%
%
\caption{(a-e) Evolution of the $\sigma^-$-polarized magneto-photoluminescence in vertical electrical fields between 88 and 4.2 kV cm$^{-1}$. To highlight the magnetic field effects, (f-j) show the evolution after subtraction and normalization by the spectrum at 0 T. Corresponding figures for $\sigma^+$ detection and details about the Gaussian deconvolution procedure are given in Ref.~\onlinecite{SuppMat}}
\label{fig:MPL}       
\end{figure*}

\section{Experimental details}
\label{sec:Exp}
\subsection{Sample and device}
A p-i-n diode was grown by molecular beam epitaxy (MBE) on a n-type GaAs (001)
substrate. The intrinsic GaAs region spans 400 nm embedding in its center a single layer of self-assembled InAs QDs.  After the formation of the QDs, they were covered by a
6-nm-thick GaAsSb layer with 28\si{\percent} nominal content of Sb. More details about the growth recipe and the morphological changes induced by rapid thermal annealing (RTA) treatment of these QDs can be found elsewhere.~\cite{ulloa_high_2012} After the RTA treatment, mesas of different sizes and ohmic contacts were defined by conventional
optical lithography techniques.

\subsection{Optical characterization}
To investigate OABE in our sample, magneto-photoluminescence
(MPL) spectra were recorded at 5 K in the Faraday configuration up to 9 T. 
To avoid unnecessary manipulations, the light polarization was analyzed in
the circular basis reversing the magnetic field direction. After the circular polarization analyzer, the emission was coupled into a multimode optical fibre and then imaged into the 100-$\mu$m-wide slits of a 300 mm focal length monochromator (1200 lines/mm grating). The light intensity was measured with a peltier cooled InGaAs photomultiplier connected to a lock-in amplifier. For every magnetic field, a PL spectrum was recorded for $V
(\textrm{V})=-2$,$-$1, 0, 1 and 1.35 (corresponding to 88, 63, 38, 13 and 4.2 kV cm$^{-1}$,
respectively). $V$ $<$ 0 ($>$ 0) corresponds to reverse (forward) bias as indicated in Fig.~\ref{fig:Intro} The power of the temperature stabilized 690 nm diode laser was registered simultaneously being its value constant within $\pm 4$\% during the whole experiment and within $\pm$0.2\% while scanning the voltage at fixed magnetic field. 
Each spectrum was normalized by its corresponding excitation
power and then analyzed by Gaussian deconvolution.

\section{Bias Dependent Magneto-Photoluminescence Experiments}
\label{sec:VBz}

Figure~\ref{fig:MPL}(a-e) shows the evolution of the $\sigma^-$-polarized MPL spectra in vertical electrical fields ($F_z$) between 88 and 4.2 kV cm$^{-1}$. To highlight the magnetic field effects, the lower panels of Figure~\ref{fig:MPL} show the evolution after subtraction and normalization by the spectrum at 0 T. In our previous work, we investigated the electric field response at 0 T of the type-II InAs/GaAsSb QD system.~\cite{Llorens2015} A large reverse bias increases the electron tunneling rate out of the InAs quantum dot potential towards the GaAs barrier. This leads to a noticeable intensity quenching moving to the left in the upper panel of Figure~\ref{fig:MPL}. The electric field also reduces the electron-hole overlap and red shifts the peak emission energy at 0 T. In the following, we focus on the changes produced by $F_z$ in the magnetic response of the sample. To maintain an unified description, the  excitation conditions were kept approximately the same in the experiments of Ref.~\onlinecite{Llorens2015} and here. Thus, the PL emission comprises two bands which are split by $\approx$36 meV at 0 V and 0 T. These bands arise from the ground state and bright excited state recombinations, both inhomogeneously broadened. Their integrated intensity ratio varies with $F_z$ and is always larger than 7.

The magnetic properties will be discussed attending solely to the evolution of the PL near the ground state of the system. For every $F_z$, the $\sigma^+$ and $\sigma^-$ MPL spectra were recorded up to 9 T. Gaussian deconvolution was then applied to extract the overall evolution of the ground state emission. From the deconvoluted MPL peak energy, $E_{\sigma}$, and integrated intensity, $I_{\sigma}$, we obtain the average energy shift,
$E_{\alpha}=(E_{\sigma^+}+E_{\sigma^-})/2$, unpolarized integrated intensity,
$I=(I_{\sigma^+}+I_{\sigma^-})/2$, Zeeman splitting, $\Delta
E_z=E_{\sigma^+}-E_{\sigma^-}$,  and degree of circular polarization,
$DCP=(I_{\sigma^+}-I_{\sigma^-})/(I_{\sigma^+}+I_{\sigma^-})$
for the low energy band alone. 

\begin{figure}[t]
\includegraphics[width=\columnwidth]{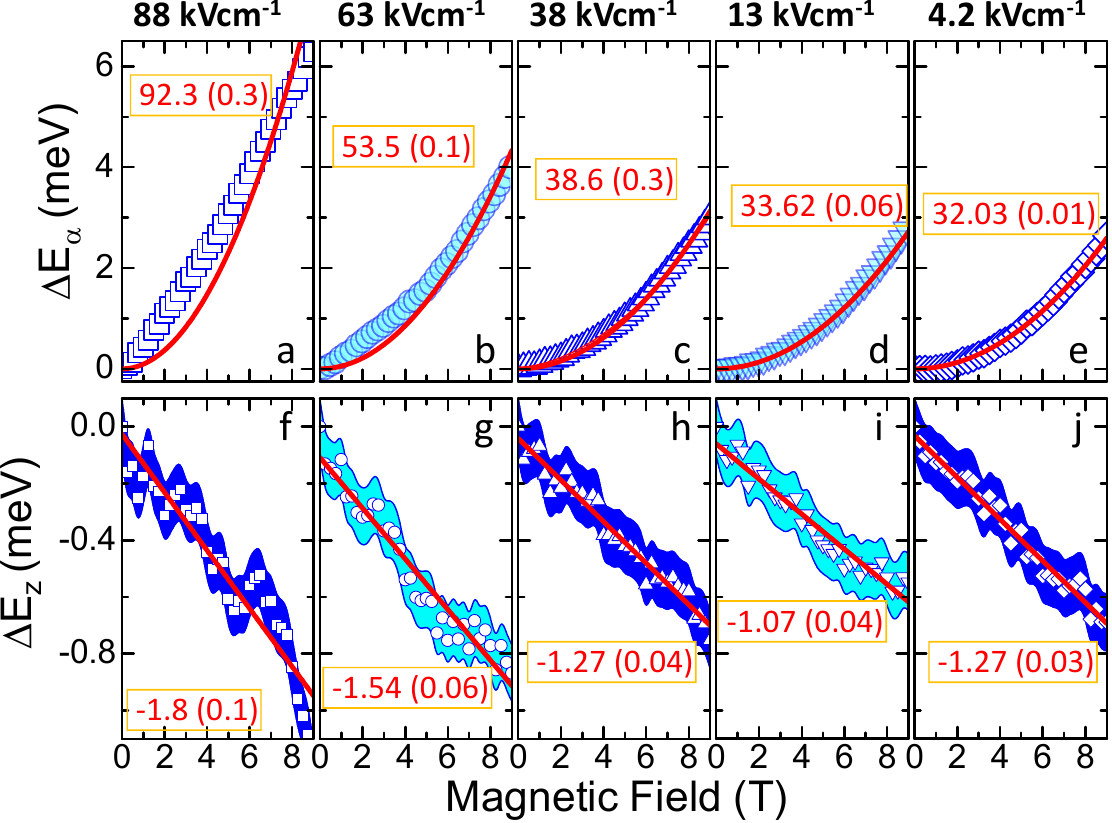}
%
%
\caption{(a-e) Dependence of the diamagnetic shift on the applied electric field. Solid lines stand for parabolic fits obtained with the indicated diamagnetic shift coefficients, $\alpha_D$. (f-j) Experimental Zeeman splittings (symbols) and corresponding $g$-factors extracted from their linear dependence (solid lines). Shaded areas correspond to the standard errors arising from gaussian deconvolution of the emission band.}
\label{fig:MPL_resE}       
\end{figure}

Figure~\ref{fig:MPL_resE} gathers the diamagnetic energy shift $\Delta
E_{\alpha}=E_{\alpha}(B)-E(0)$ and Zeeman splitting for different electric
fields. Solid lines are quadratic and linear fittings to $\Delta
E_{\alpha}(B)=\alpha_D B^2$ and $\Delta E_z (B)=\mu_B g B$, respectively, where
$\mu_B$ is the Bohr magneton, and $\alpha_D$ and $g$ represent the diamagnetic
shift coefficient and Land\'{e} $g$-factor for the type-II exciton,
respectively.~\cite{HayneLuminescence_27_179_2012} Both functions describe
accurately the data, except for a linear deviation for $\Delta E_{\alpha}$ in
the strong tunneling regime corresponding to relatively large electric fields 
in Figs.~\ref{fig:MPL_resE}(a-b). We observe that both
$\alpha_D$ and $g$ are affected by the external electric field and also that
oscillations in magnetic field are absent for the ground state peak energy. In
this bias range, the $g$-factor variation is $\approx 80$\si{\percent} and
shall be attributed to a voltage modulation of the spin-orbit interaction as it
will be discussed in detail in Section~\ref{sec:SOC}. 

%

The $\alpha_D$ variation is even more pronounced and exhibits a clear diminishing trend
reducing the reverse bias. For a magnetic field applied in the growth
direction, the diamagnetic shift is usually associated with the exciton
wave function extension in the growth plane.~\cite{nash_diamagnetism_1989}
However, this simple correlation between exciton diameter and diamagnetic shift
magnitude is lost when electronic levels with different angular momentum cross
each other in the fundamental state. These level crossings are the source of the  interference between different quantum paths leading to AB magneto-oscillations in quantum rings.~\cite{Fomin_QR_Book_1st_eddition} In an ensemble experiment, they might be obscured by inhomogeneous effect but might reveal themselves as a flattening of
$\Delta E_{\alpha}(B)$ and the observed evolution of $\alpha_D$ \textit{vs.}
$F_z$. In our case, the tunneling also becomes more important as the
positive field increases. In these conditions, the electron wave function penetrates in the barrier and might also affect the value of $\alpha_D$. These issues shall be discussed in detail in the next sections.

The evolution of the unpolarized integrated intensity after subtraction and normalization by the 0 T value, $\Delta I$, and the
$DCP$ are depicted in Figure~\ref{fig:MPL_resI}. At large electric fields, $\Delta I$ increases with $B$ exhibiting a strong magnetic brightening up to 75\% [Figure~\ref{fig:MPL_resI}(a)].
The dependence follows a fixed linear slope with small oscillations which are only apparent in the
first derivative of the data (not shown). By reducing $F_z$, the magnetic brightening effect diminishes and, below 13 kV cm$^{-1}$, an oscillatory behavior develops [Figure~\ref{fig:MPL_resI}(b)].
In the same bias range, the $DCP$ is positive, as
expected from the negative $g$-factor, and evolves with magnetic field from a
rather monotonic dependence at 83 kV cm$^{-1}$ to a clearly oscillatory one at
4.2 kV cm$^{-1}$.

\section{Discussion}
\label{sec:Discusion}

\begin{figure}[t]
\includegraphics[width=0.70\columnwidth]{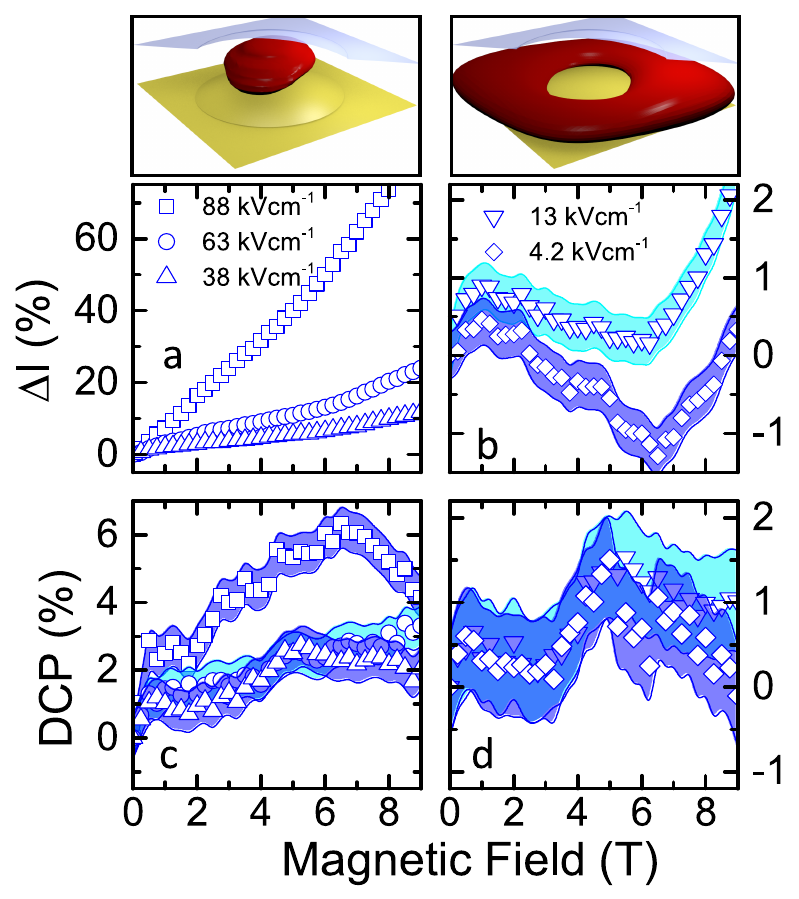}
%
%
\caption{Evolution of the unpolarized integrated intensity and degree of circular polarization in the high (a, c) and low (b, d) bias regime. The top panels represent the qualitative evolution of the hole probability density between both regimes. Shaded areas correspond to the standard errors arising from gaussian deconvolution of the emission band.}
\label{fig:MPL_resI}       
\end{figure}

In the quantum ring literature, MPL intensity oscillations are related to 
changes in the ground state angular momentum arising from the OABE.~\cite{Fomin_QR_Book_1st_eddition}
Voltage modulation of such oscillations have been reported for type-I InAs/GaAs
single quantum rings,~\cite{ding_gate_2010,li_tunable_2011} and also predicted for 2D materials.~\cite{de_sousa_unusual_2017} The modulation
must be associated to changes in the effective quantum ring confinement
potential. In the type-II InAs/GaAsSb QD system, holes are spatially localized within the GaAsSb layer near the InAs QDs.~\cite{Llorens2015,ulloa_gaassb-capped_2010,
ulloa_analysis_2012,klenovsky_electronic_2010,hospodkova_type_2013,
liu_long-wavelength_2005} The strain distribution and electron-hole Coulomb interaction create a net attractive potential for the hole that is balanced by the InAs/GaAsSb and GaAs/GaAsSb valence band offsets. An external electric field modulates the geometry of this potential and, in turn, a change of the hole wave function topology from singly to doubly connected is expected. The inset in Figure~\ref{fig:MPL_resI} represents the qualitative evolution of the hole probability density according to the theoretical description developed on Ref.~\onlinecite{Llorens2015}.

%

\subsection{Axially symmetric $\vec{k}\cdot\vec{p}$ model}
\label{subsec:kpEzBz}

\begin{figure*}[t]
        \includegraphics[width=0.90\textwidth]{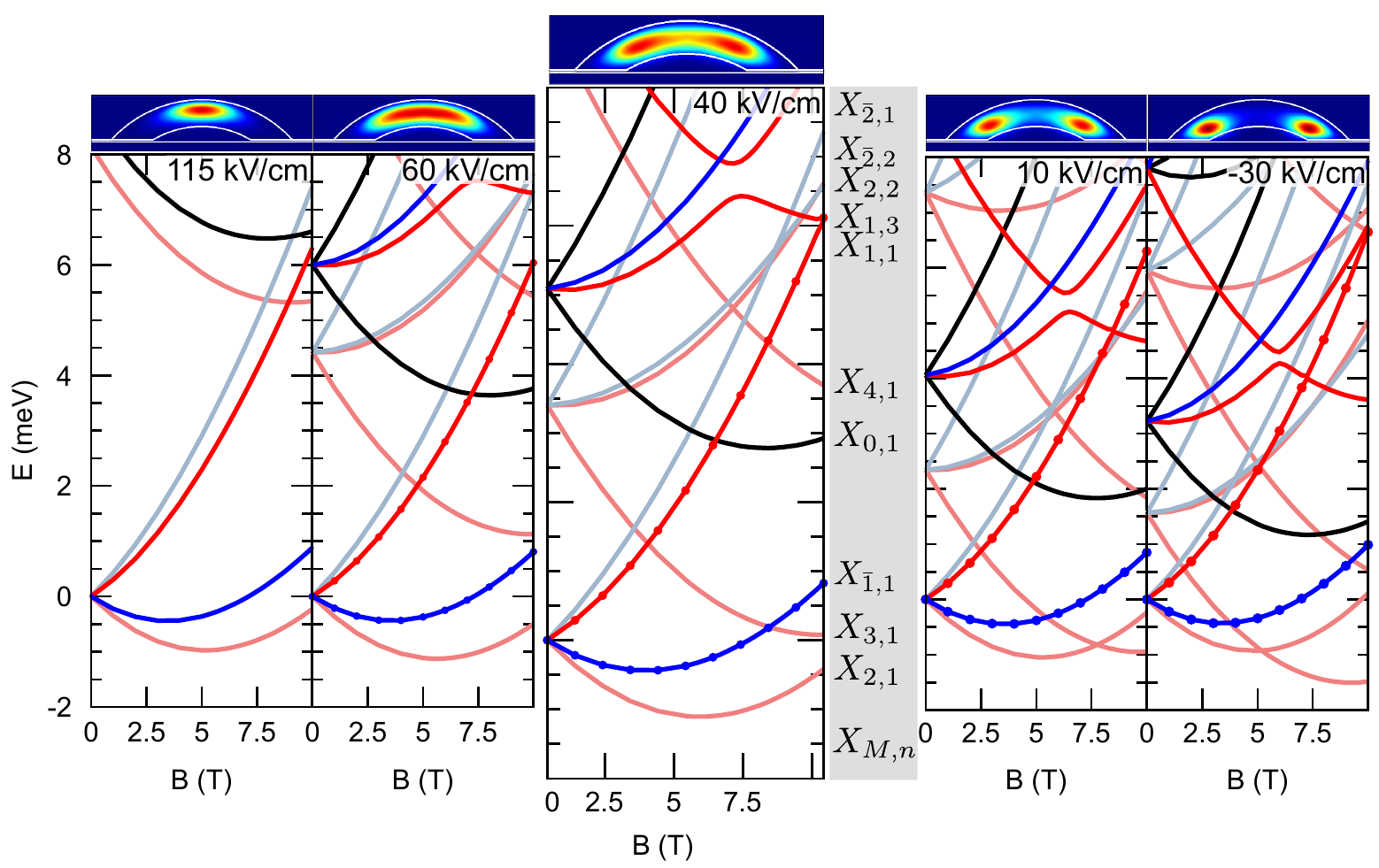}
%
%
\caption{Exciton energy levels relative to the ground state energy at $B=0$ T.
        Blue lines are states with $M=-1$, red lines $M=1$, light-red
        lines $M>1$, light-blue $M<-1$ and black $M=0$. The spot
        size is proportional to the radiative rate and is brought to the same scale
        throughout the pannels. The contour plot on top of
        each plot depicts the probability density of the hole ground state at
        $B=0$ T. The axes dimensions are 50 nm and 12.5 nm.}
\label{FIG:E_Bz_F}       
\end{figure*}

To analyze our data quantitatively, we start with an axially symmetric model
for the type-II InAs/GaAsSb QD system. Under this approximation, the total
angular momentum is a good quantum number for the exciton states and allows a
proper labeling scheme.  The whole $\vec{k}\cdot\vec{p}$ Hamiltonian including
the strain is forced to be axially symmetric. Hence, even in the multiband
case, the states are labeled according to the $z$-projection of the total
angular momentum ($M$). In case of small band mixing, the state can also be
labeled with the $z$-projection of the orbital angular momentum of the envelope
function associated with the main Bloch component ($m$). Further
details and explicit expressions can be found in the Appendix~\ref{App:KP}.

The numerical results are obtained for a quantum dot of the same size and
composition as those analyzed in our previous
studies~\cite{ulloa_high_2012,Llorens2015}. The geometry of the InGaAs QD is a
lens shape of radius $R_\text{QD}=11$ nm and height $H_\text{QD}=3$ nm.  It
sits on top of a 0.5 nm InGaAs wetting layer (WL) and is surrounded by a GaAsSb
shell of 6 nm thickness which plays the role of the overlayer. The composition
of the nanostructure is Ga$_{0.25}$In$_{0.75}$As and that of the overlayer
GaAs$_{0.80}$Sb$_{0.20}$.

Figure~\ref{FIG:E_Bz_F}  shows the results in the exciton picture. Each
electron (hole) state is defined by the quantum number $M_{e(h)}$ introduced in
Eq.~\eqref{EQ:Psi}. Hence, the exciton
states are defined by the addition of the total angular momentum of the
constitutive particles $M = M_\text{e} + M_\text{h}$ ($z$-projections). The
only optically active exciton states are those characterized by $M=0, \pm1$.
The emitted photons are polarized along the $z$ direction for $M=0$ and
circularly polarized $\sigma_\pm$ for $M=\pm1$. We have selected five values of
the electric field which fully cover the topological transition from dot to
ring.  For clarity, we have added a contour plot of the hole probability
density on top of each energy level dispersion. One can easily distinguish the
localization of the hole particle on top of the nanostructure (115 kV
cm$^{-1}$) and its drift towards the base as the electric field diminishes. The
corresponding exciton states are indicated only for the central panel as
$X_{M,n}$, being $M=0,\pm1,\pm2,\ldots$ the exciton total angular momentum
$z$-projection and $n=1,2,\ldots$ the principal quantum number of that
particular $M$ manifold.  We first focus on the dispersion for $F_z=115$ kV
cm$^{-1}$ (large reverse bias).  At $B=0$ T, the ground state is composed of
the fourth-fold degenerated quadruplet with $n=1$ and $M=\pm1$ and $\pm2$.  The
magnetic field splits this level into four branches. The $M=\pm1$ can be
distinguished by the non-zero radiative rate (dots over the line) and more
intense colored lines used in the representation. These four exciton states
result of the combination of two electron and two hole states whose dominant
Bloch amplitudes are $\ket{1/2, \pm1/2}$ and $\ket{3/2, \pm3/2}$. The envelope
function of each dominant Bloch amplitudes is defined by $m\approx0$.  This
explains the weak dependence of the internal structure of the quadruplet to the
electric field and hence to the topology of the hole wave function.
Irrespective of whether the hole is located above the apex of the QD ($F=115$
kV cm$^{-1}$) or close to its base ($F=-30$ kV cm$^{-1}$) the energy levels are
weakly perturbed. On the contrary, the quantization energy of the excited
states is significantly perturbed by the electric field. The first excited
state at $B=0$ T is out the plot for $F=115$ kV cm$^{-1}$, while at $F=-30$ kV
cm$^{-1}$ the energy splitting with the ground state is $\approx1.5$ meV. This
clearly illustrates how the electric field efficiently modifies the angular and
radial confinement exerted on the hole wave function. In theses states, the
dominant envelope wave functions of the hole is characterized by $|m|>0$ and
therefore are more sensitive to the vertical position within the overlayer.

Figure~\ref{FIG:E_Bz_F} illustrates that the applied electric field can effectively
isolate the ground state from the excited states for $F>60$ kV cm$^{-1}$ or
bunch excited and ground states together for $F<60$ kV cm$^{-1}$. In the bunching regime, the model predicts the crossing between states of different $M$, and thus a change in the ground state orbital confinement. More explicitly, the $X_{3,1}$
high energy state and the $X_{\bar{1},1}$ and $X_{2,1}$ low energy states cross at $B=5.4$ and 7.8 T for $F=10$ kV cm$^{-1}$,
respectively.  These crossings occur at fields within the range of the
experimental values presented before. This would result in the possibility
of observing experimentally, at least, the crossing between $X_{3,1}$ and the
states of lower energy and thus an associated optical AB oscillation.
The precise correspondence with the experimental
results is difficult to establish given that its observation would depend on
the thermalization of the excited carriers and the eventual relaxation of the
selection rules in the actual QDs, as explained below. As expected for AB
related effects, the number of
crossings and their actual positions depend on the exciton in-plane dipole
moment which, in our case, decreases with the electric field. In view of these results, the oscillatory
dependence observed for the unpolarized integrated intensity and $\mathit{DCP}$ in Fig.~\ref{fig:MPL_resI} is a consequence of a change in the total angular momentum of the ground state and hence a signature of the OABE.

\begin{figure}[t]
        \includegraphics[width=0.75\columnwidth]{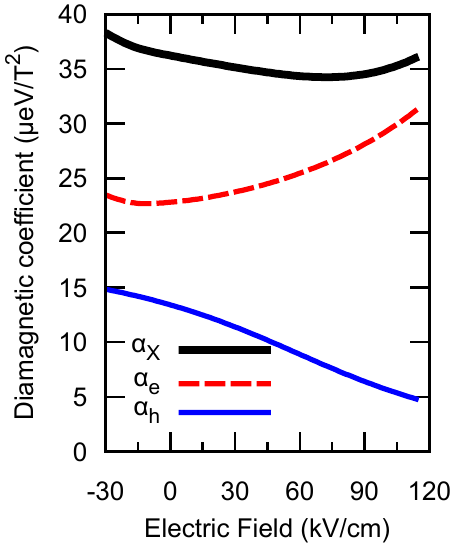}
%
%
        \caption{Diamagnetic coefficient for the electron 
        (dashed red), hole (solid blue), and the exciton (thick-solid black) states.
}
\label{FIG:ALPHA_X}       
\end{figure}

The inhomogeneous
broadening of the MPL spectra in Sec.~\ref{sec:VBz} prevents the observation of the individual energy levels 
shown in Fig.~\ref{FIG:E_Bz_F}, thus peak energy magneto-oscillations cannot be detected in our experiment.~\cite{kim_observation_2016} This notwithstanding, our analysis of the emission energy shifts in Figures~\ref{fig:MPL_resE}(a-e) revealed a three fold increase of the diamagnetic shift coefficient with the electric field. From the numerical results shown in Figure~\ref{FIG:E_Bz_F}, we can calculate this magnitude for the bright exciton 
states. The resulting coefficients
for the electron $\alpha_e$, hole $\alpha_h$ and exciton $\alpha_X$ are shown
independently in Figure~\ref{FIG:ALPHA_X}. 

The hole
diamagnetic shift decreases from
$\sim$15 to $\sim$5 \si{\micro eV/T^2} by increasing the electric field from -30 to 115 kV cm$^{-1}$. This is expected since, during the hole drift from the base towards the apex of the overlayer, its density of probability shrinks decreasing $\alpha_h$
(see contour plot insets in Fig.~\ref{FIG:E_Bz_F}).~\cite{nash_diamagnetism_1989} In the same bias range, $\alpha_e$ goes from from $\sim$23 to $\sim$32 \si{\micro eV/T^2},
increasing with the reverse bias. This is a consequence of the spill-over of
the electron wave function in the GaAs barrier underneath. 
For $F_z>60$ kV cm$^{-1}$, electron tunneling out of the
InAs QD becomes noticeable.~\cite{Llorens2015} The numerical model catches this situation only approximately and, for the largest bias, the calculated $\alpha_X$ is only 37 \si{\micro eV/T^2}, while experimentally $\alpha_D$ is 92.3 \si{\micro eV/T^2}. Indeed, for $F_z>63$ kV cm$^{-1}$, the experimental dispersion becomes increasingly linear. Linear dispersion is characteristic of the Landau regime. In our context, it corresponds to a weaker confinement as a result of the delocalization of the electron wave function as the tunnel regime becomes dominant [see Figs.~\ref{fig:MPL_resE}(a,b)].  In the opposite bias regime, the correspondence with the experimental values
is much better. $\alpha_D$ (32.62 \si{\micro eV/T^2}) is found at $F=4.2$ kV cm$^{-1}$ while $\alpha_X$ renders 36.05 \si{\micro eV/T^2}. As shown in Fig.~\ref{FIG:E_Bz_F}, in this bias range, the ground state of the electron-hole system is composed mostly of states with angular momentum $|M|>1$. Under the axially symmetric approximation, these states are purely dark and do not contribute to $\alpha_X$. If the symmetry of the actual confinement potential is broken, they can achieve oscillator strength, lower further the diamagnetic shift and produce AB intensity oscillation, as discussed in the next Section.

\subsection{In plane asymmetry effects}
\label{subsec:asymmetry}

It has been profusely reported that quantum dot
elongation can take place after capping. The strain fields that build up during
this process provoke the anisotropic segregation of In atoms leading to an
eccentricity increase of previously cylindrical
systems.~\cite{alonso-alvarez_strain_2013,teodoro_-plane_2012} Analogous
process is also triggered during the synthesis of type-II InAs/GaAsSb quantum
dots which also might lead to anisotropic piezoelectric
fields.~\cite{klenovsky_electronic_2010,krapek_excitonic_2015} Thus, it is a
goal of our following discussion to assess the effects of in-plane confinement
anisotropy in the electronic properties. To this end, we introduce an effective
mass description of both the conduction and valence bands, and incorporate the
effects of confinement asymmetry for electrons and holes in a model that can
emulate quantum dots and rings within the same framework, as well as the
resulting Rashba spin-orbit coupling fields arising from confinement and
external fields.

\begin{figure}[t]
\includegraphics[width=\columnwidth]{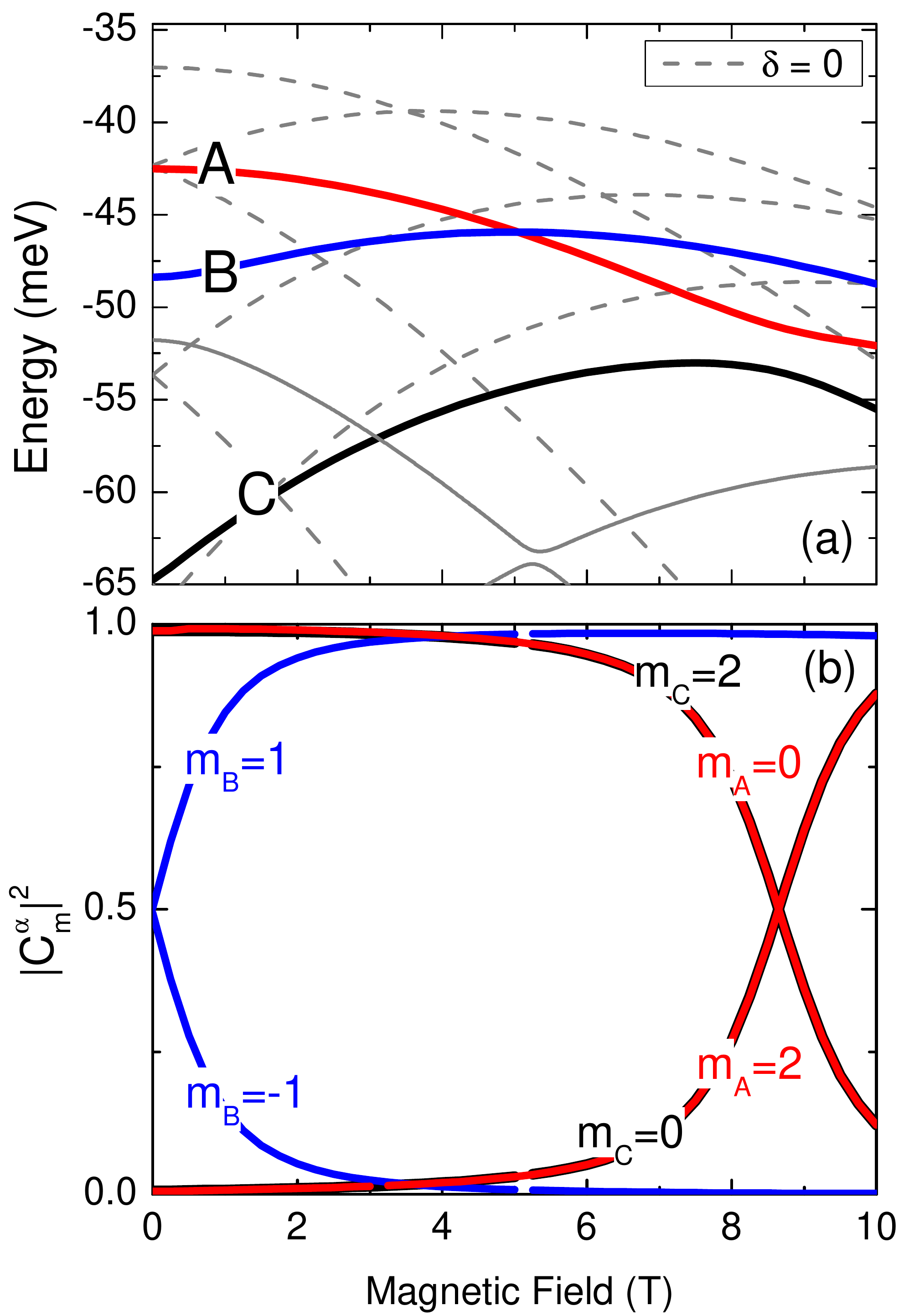}
\caption{(a) Calculated energy levels in a quantum ring for the top valence subband: solid curves correspond to an eccentric ring with $\delta=2$ meV/100 nm$^2$ while the dashed curves are calculated for a cylindrical geometry with $\delta=0$. (b) Calculated weight coefficients of the A, B and C energy levels in (a) as a function of the magnetic field strength. (c) Calculated oscillator strength for the interband transitions involving the conduction band ground state and the admixed valence band levels A and C.}
\label{fig1}       
\end{figure}

The eigen-value problem for the conduction band will be solved by expanding the
corresponding wave functions in the basis of the eigen-solutions of the
following effective mass Hamiltonian,
\begin{equation}\label{Hamiltonian}
H=\frac{\hbar^2}{2\mu^*}k^2+V_0 (z)+V_0 (\rho)+\frac{g^* \mu_B}{2}\mathbf{B}\cdot \bm{\sigma}~,
\end{equation}
with $\mu^*$ the effective mass, $\mathbf{k}=-i \bm{\nabla} +e/\hbar \mathbf{A}$ and the magnetic field pointing
along the growth $z$-direction, where $\bm{\sigma}$ is the Pauli matrix
vector, and $\mathbf{A}=B/2 \rho \widehat{\varphi}$. For the unperturbed basis,
$V(z)$ will be assumed as a rigid wall potential profile while the in-plane
confinement in polar coordinates takes the form:~\cite{tan}
\begin{equation}\label{inplanepotential}
V_0(\rho)=\frac{a_1}{\rho^2}+a_{2}\rho^2-2\sqrt{a_{1}a_2},
\end{equation}
which allows obtaining an exact solution that covers both the quantum ring
and the quantum dot confinements. The parameters $a_{1}$ and $a_{2}$ define the
structure shape and for $a_{1}\neq 0$, a ring with radius
$R_\text{QR}=(a_{1}/a_{2})^{1/4}$ is obtained. In turn, by setting $a_{1}=0$, a
parabolic quantum dot with effective radius
$R_\text{QD}^2=\hbar/(2\pi\sqrt{2a_2{\mu^{*}}})$ can be emulated. 
This potential allows us to describe effectively the change of topology
on the hole as a result of the applied electric field.
             
For the valence band basis, used to expand the Luttinger Hamiltonian
eigen-solutions, we use an analogous separable problem yet assuming
anisotropic effective masses in Eq.~\eqref{Hamiltonian}, and replacing $g^*/2 $
and $\sigma_z$, by $-2 \kappa$ and the angular momentum matrix for $j=3/2$,
$J_z$, respectively.

The symmetry constrains can be subsequently relaxed by reshaping the in-plane
confinement in the following way,
$V(\rho,\varphi)=V_0(\rho)+\delta\cdot \rho^2\cos^2{\varphi}$. This is
an extension of the profile proposed in Refs.~\onlinecite{2,VivaldoSergio}, where the
term controlled by the parameter $\delta$ determines the eccentricity of the
outer rim of the confinement. The values of the eccentricity are given by
$e=\sqrt{1-a_{2}/(a_{2}+\delta)}$, for $\delta>0$, that corresponds to an
elliptical shrinking, or $e=\sqrt{1-(a_{2}+\delta)/a_{2}}$ for $\delta<0$, that
leads to an elliptical stretching. Additionally, an electric field along the
growth direction can be considered by inserting the term $e F z$ into $V(z)$
that will couple the wave function components of different parity. The details
on the solution of the Schr\"{o}dinger equation are in Appendix \ref{App:EMA}. The asymmetric solution is given by the expansion
\begin{equation}\label{wave function}
\Psi=\sum_{\alpha,n,m,l} C_{n,m,l}^{\alpha} \psi_{n,m,l}^{\alpha}(\rho,\varphi,z),
\end{equation}
where $\alpha$ labels the basis functions at the Brillouin zone center in the Kane model ($|1/2,\pm1/2\rangle$, $| 3/2,\pm 3/2 \rangle$ and $| 3/2, \pm1/2\rangle$), $m$ is the $z$-projection of the orbital angular momentum, $n$ is the radial quantum number and $l$ is the vertical quantum number. Thus, the expansion coefficients, $C_{n,m,l}^{\alpha}$, determine the spin character of the final state, namely, the degree of hybridization of the various $m$ components.

The eccentricity breaks the cylindrical symmetry. Figure~\ref{fig1}(a) shows the impact of the symmetry break on the valence band energy levels for the first few states confined in a QR with $R_\text{QR}=12$ nm and height $h_\text{QR}=7$ nm. For this calculation, we have omitted the spin splitting. Dashed lines render the cylindrical case ($\delta=0$ meV) mimicking the energy dispersion found in the previous section.
Solid lines stand for an eccentric ring with $\delta=2$ meV. The only relevant energy levels for the discussion
are represented with bold lines and labeled as A, B and C. The A and B levels cross at $B$=4 T, and the A and C levels anti-cross at $B$=7.5 T. The mixing of different $m$-components reduces the effective diamagnetic shift of the valence band ground hybrid state. This mixing is quantified in Fig.~\ref{fig1}(b), where the expansion coefficients $\left|C_{0,m,1}^{\alpha}\right|^2$ resulting from the diagonalization of the valence band Hamiltonian have been depicted. In the symmetric case, the ground state experiences a sharp change at the crossing from a $m=0$ character to a $m=1$ (not shown). In contrast, in the eccentric case there are no discontinuities, reflecting a soft evolution of the wave function character with the magnetic field. The states corresponding to the energy levels A and C swap their character from $m=0$ to $m=2$ at the anti-crossing. The energy level B acquires a dominant $m=1$ character being insensitive to the crossing and anti-crossing. 

\begin{figure}[t]
\includegraphics[width=\columnwidth]{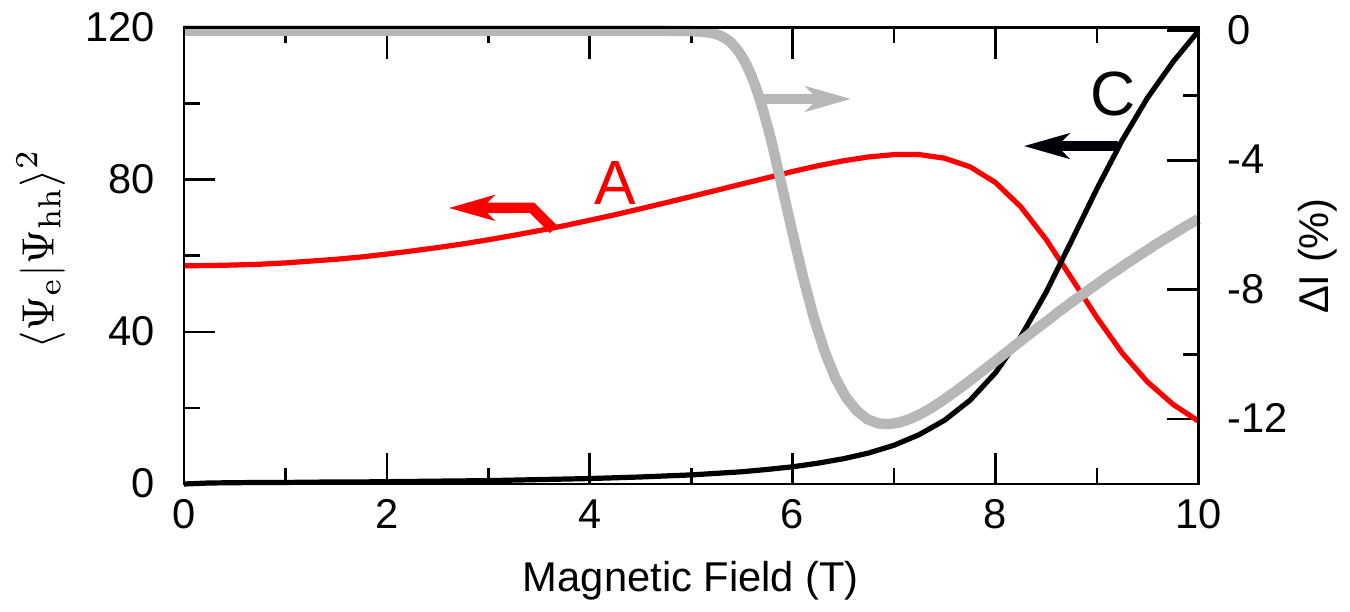}
\caption{Calculated oscillator strength for the interband transitions involving the conduction band ground state and the admixed valence band levels A and C (thin red and black lines, respectively). Computed relative intensity $\Delta I$ with a rate equations model (thick grey line). The electrons are calculated for a QD of $R_\text{QD}$=15 nm and $h_\text{QD}$=7 nm.
The eccentricity is $e$=0.53.}
\label{fig8}       
\end{figure}

The changes of the electronic states energy and character modulate also the emission intensity, which can be calculated from the oscillator strength (OS). As the electron ground state has $m$=0 character, the state B is optically inactive within the model, being its OS negligible. States A and C, however, show an exchange of oscillator strength as they approach the anti-crossing as shown in Figure~\ref{fig8}. The smooth increase of the OS at high magnetic fields is attributed to the magnetic brightening. In this situation, we can calculate the emission intensity of the system solving a simple rate equations model based on Ref.~\onlinecite{sugawara_quantumdot_2002} as explained in Ref.~\onlinecite{SuppMat}. The radiative rates are parameters of the model and scale with the OS calculated for the states A and C. For state B, the pure dark condition is relaxed, since dark states must eventually recombine due to hybridization processes not considered here (e.g. electron-hole exchange). As shown in Figure~\ref{fig8}, our model predicts an intensity oscillation around 7 T with similar phase but slightly larger amplitude than the experimental evolution displayed in Figure~\ref{fig:MPL_resI}(b). It must be noted that the position of the crossings and the energy difference between states presents fluctuations among QDs. This shall lead to a shallower experimental oscillation resulting from the ensemble average of many curves like the ones shown in Figure~\ref{fig8}.~\cite{Marcio_AB_2010}


\subsection{Spin-orbit coupling effects}
\label{sec:SOC}

We have just shown that the symmetry reduction in a quantum ring lowers the effective diamagnetic shift of the ground state as the angular momentum states become intermixed. In turn, including spin-orbit interaction terms, the electrical modulation of the lateral and vertical confinement also changes the spin states.~\cite{ares_nature_2013,bennett_voltage_2013,corfdir_tuning_2014,jovanov_observation_2011,klotz_observation_2010,prechtel_electrically_2015,yang_tuning_2015,tholen_strain-induced_2016} According to the results presented in Fig.~\ref{fig:MPL_resE}(f-j), the exciton energy spin
splitting increases with the device reverse bias. Meanwhile, within the GaAsSb overlayer, the hole wave function geometry and topology evolve. Both observations can be connected once the spin orbit interaction is incorporated to the
model represented by Eqs.~\eqref{Hamiltonian}-\eqref{wave function}. The interaction is introduced through the
Rashba contribution to the total Hamiltonian $H_R^c=\alpha_c \bm{\sigma}\cdot
(\nabla V \times \mathbf{k})$, for the conduction band, or $H_R^v=\alpha_v
\mathbf{J}\cdot (\nabla V \times \mathbf{k})$, for the valence band.
The expression for the spin orbit Hamiltonian assuming the confinement profile that includes all the asymmetry terms and the resulting energy corrections are given in Appendix \ref{App:SOC}. 

In turn, we can separate the spin-orbit effects induced by confinement and
asymmetry into first or higher order contributions. The latter ones are
produced by the coupling of previously unperturbed levels and appear strongly
when these levels approach inducing (or enhancing) anticrossings, mostly at higher
fields. The first order terms appear already at vanishing fields and may
provoke, for instance, the tuning of the effective Land\'{e} factor. We shall focus
the discussion on these ones.

\begin{figure}[t]
\includegraphics[scale=1.0]{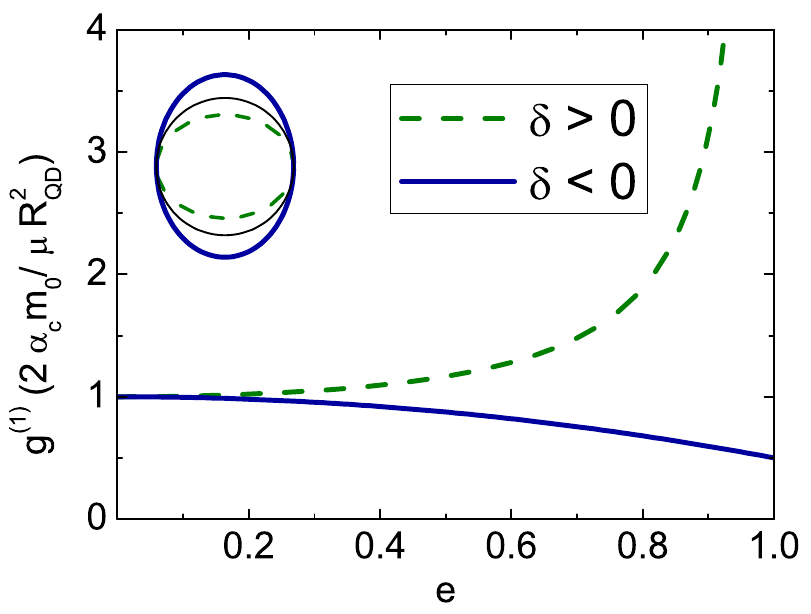}
\caption{Calculated first order correction of the conduction band Land\'{e} factor as function of the eccentricity for a quantum dot according to Eq.~\eqref{g1}}
\label{figSOBC}       
\end{figure}

The first order correction to the conduction band Zeeman splitting induced by the spin-orbit coupling can be calculated exactly for the ground state of a quantum dot according to:~\cite{Cabral}
\begin{equation}\label{g1}
g_c^{(1)}=\frac{2 m_0 \alpha_{c}}{\mu^* R_\text{QD}^2} \left[   1+ \text{sign}(\delta) \frac{e^2}{2- \left[1+ \text{sign}(\delta)\right]e^2}       \right].
\end{equation}
This result is depicted in Fig.~\ref{figSOBC} and illustrates how the Land\'{e} factor
correction grows as the quantum dot volume is reduced by shortening
$R_\text{QD}$ or by shrinking the lateral confinement into an ellipse ($\delta>0$ corresponds to an elliptical shrinking). 

\begin{figure}[t]
\includegraphics[scale=1.0]{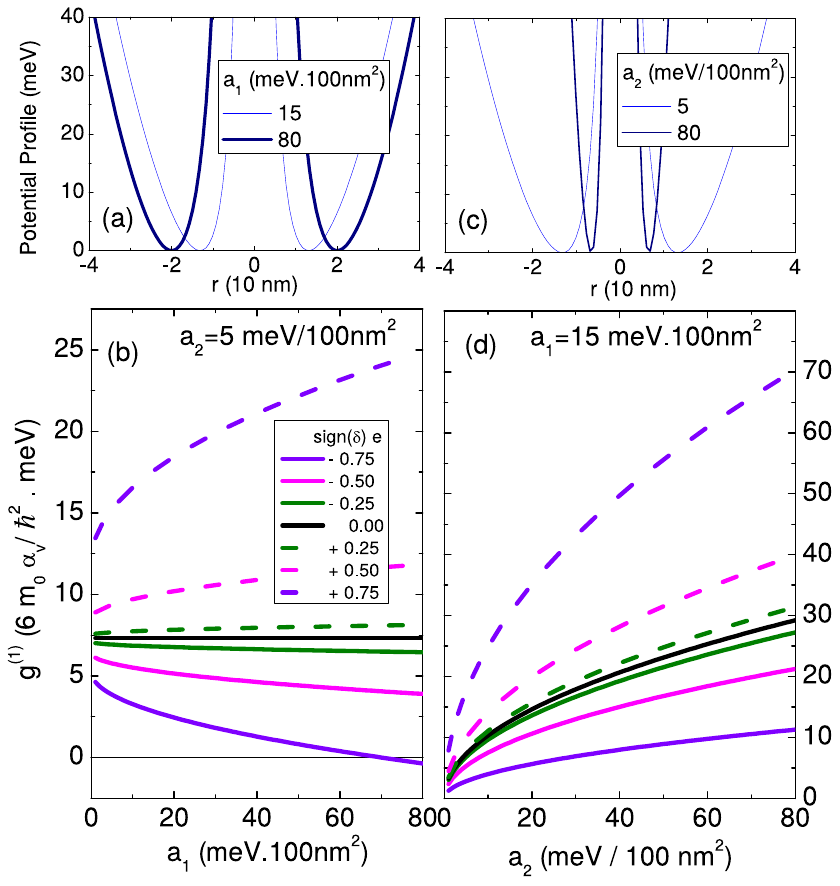}
\caption{(a) In-plane confinement profiles for fixed $a_2$ and varying $a_1$. (b) Calculated first order correction of the Land\'{e} factor for fixed $a_2$ and varying eccentricity as function of $a_1$.(c) In-plane confinement profiles for fixed $a_1$ and varying $a_2$. (d) Calculated first order correction of the Land\'{e} factor for fixed $a_1$ and varying eccentricity as function of $a_2$.}
\label{figSOBV}       
\end{figure}

For a quantum ring, such a modulation is not trivial and depends on the way the confinement shape is modified as shown in Fig.~\ref{figSOBV}. Note that by
fixing $a_2$ and increasing $a_1$, the quantum ring radius increases as a result
of widening the inner rim [Fig.~\ref{figSOBV}(a)]. Such a modulation of the confinement profile can either increase or decrease the valence band Land\'{e} factor correction according to the value of the eccentricity and the sign of $\delta$ as displayed in Fig.~\ref{figSOBV}(b). For large enough rings, the eccentricity may even lead to an
absolute reduction of the Land\'{e} factor. If in contrast, the parameter $a_1$ is fixed while increasing $a_2$, the confinement potential shrinks as a result of the reduction of the external radius. This condition is plotted in Fig.~\ref{figSOBV}(c) while the corresponding first order correction to the $g$-factor appears in panel (d). Such a reduction of the quantum ring radius provokes a monotonic increase of this correction factor regardless the value of the eccentricity.

Through this model, the non-monotonic increase of the $g$-factor displayed in Fig.~\ref{fig:MPL_resE} can be ascribed to the bias dependent shrinkage and symmetry reduction of the hole wave function. To assess the absolute values of the $g$-factor correction, and the relevance
of antimonides in that respect, we may contrast systems with different
spin-orbit coupling: GaAs, GaSb and InSb, respectively. For the valence band of these materials, $\alpha_v^\text{GaAs}=-14.62$
{\AA}$^2$, $\alpha_v^\text{GaSb}\approx 3.3\alpha_v^\text{GaAs}$,  $\alpha_v^\text{InSb}\approx 38\alpha_v^\text{GaAs}$. Their calculation is detailed in Appendix~\ref{App:SOC}.
According to these values, the units used in Fig.~\ref{figSOBV} for the valence band Land\'{e} factor corrections are -0.012, for GaAs, -0.038, for GaSb, and -0.442, for InSb.
In turn, for the conduction band Landé factor correction, plotted in Fig. \ref{figSOBC} the units are radius dependent and for a pure InAs QD with $R_\text{QD}=8$ nm they equal 1.6. 

For electron-hole pairs confined in type-II nanostructures, the unperturbed effective Zeeman splitting can be obtained from, $\Delta E_c^{(0)}-\Delta E_v^{(0)}=(g_c^0 + 6\kappa)\mu_B B$. In the case of an electron in InAs and a hole in GaAs$_{0.8}$Sb$_{0.2}$, the effective Land\'{e} factor is $g^{(0)}_{e-h}=g^{(0)}_{c}(\text{InAs}) + 6\kappa(\text{GaAs}_{0.8}\text{Sb}_{0.2})= - 3.62$. By introducing the effects of confinement, the Landé factor can be corrected as $g_{e-h}=g^{(0)}_{e-h}+g_c^{(1)}-g_v^{(1)}$. Note that according to the values of $\alpha_c>0$ and $\alpha_v<0$, the correction leads to a Landé factor increase in relative terms towards more positive values. For low fields, the electrons and holes are confined in QD shapes with $a_1=0$ and $a_2$=5 meV/100 nm$^{2}$ for holes while $R_\text{QD}$=8 nm for electrons. This leads to Land\'{e} factor corrections of $g_c^{(1)}$=1.6 and  $g_v^{(1)}$=0.129 for electrons and holes, respectively, leading to a total Land\'{e} factor $g_{e-h}$=-1.891.  At higher electric fields the hole wave function moves downwards, where we can expect a radius increase of an eccentric QR ($R_\text{QR}^2$=$\sqrt{a_1/a_2}$) with $e=0.53$. This translates into the model by a growing $a_1$ from 0 to 10 meV$\cdot$100 nm$^2$, with $a_2$=5 meV/100 nm$^2$, corresponding to the QR with $R_\text{QR}$=12 nm characterized in Fig.~\ref{fig1}. According to Fig.~\ref{figSOBV}(b), this results in an increase on the correction of $g_v^{(1)}$=0.172. In turn, the electron is confined inside of the QD and it could move upwards, with a smaller effective $R_\text{QD}$=6.8 nm with respect to the situation of larger positive fields. In Eq.~\eqref{g1}, $g_c^{(1)}$ is inversely proportional to $R_\text{QD}$ and therefore increases its value to $g_c^{(1)}$=2.21. This translates into a total Land\'{e} factor of $g_{e-h}$=-1.23. The change in the estimation of the $g$-factor between low and high fields agrees with the observed experimental change in $g$ from -1.8 to -1.3.

\section{Conclusions}
\label{sec:conclusions}

In summary, we have presented experimental evidence on the tuning of the
electronic structure of type-II InAs/GaAsSb quantum dots under vertical
electric and magnetic fields. Induced by the external bias, the drift of the
hole wave function in the soft confinement of the GaAsSb layer encompasses a
wide range of geometrical and topological changes in the effective potential.
These changes cannot be easily induced in type-I systems and provide a rich
variety of insights on the way geometry affects the electronic structure and
thus the optical response. 
The observed effects have been studied with theoretical approaches that
include electronic confinement, strain fields and spin effects on the same
footing. In particular, we report how the application of an external bias tunes
the hole confinement geometry independently of the electron. Under certain bias
conditions, the hole confinement topology changes and magnetic field
oscillations are observed. The oscillations follow the orbital quantization of
the hole state and are thus susceptible to the confinement size and
eccentricity. Although further work is needed to put these ideas at work in single InAs/GaAsSb nanostructures, this modulation of hole orbital quantization, accompanied by the tuning of the exciton $g$-factor, might pave the way for the voltage control of spin degrees of freedom as required by several quantum technologies. 

\begin{acknowledgments}
The authors gratefully acknowledge financial support from EURAMET through EMPIR program 17FUN06-SIQUST, from Spanish MINEICO through grants TEC2015-64189-C3-2-R, MAT2016-77491-C2-1-R, EUIN2017-88844 and RYC-2017-21995, from Comunidad de Madrid through grant P2018/EMT-4308, and from CSIC through grants I-COOP-2017-COOPB20320 and PTI-001. Support from Brazilian agencies is also acknowledged through FAPESP grant 2014/02112-3 and CNPq grant 306414/2015-5.
\end{acknowledgments}

        



\appendix
\section{Details of the multi-band axisymmetric model}
\label{App:KP}
The electronic structure of the InAs/GaAsSb QDs is computed by the
$\vec{k}\cdot\vec{p}$ method folowing Trebin \emph{et
al.}~\cite{trebin_quantum_1979, winkler_book} notation.  The final Hamiltonian
is the result of adding three contribution, the $\vec{k}\cdot\vec{p}$ term
($H_{\vec{k}\cdot\vec{p}}$), the strain interactions (Bir-Pikus
$H_\varepsilon$) and the magnetic interactions ($H_B$):
\begin{equation}
H=H_{\vec{k}\cdot\vec{p}} + H_\varepsilon + H_B.  \label{EQ:hamiltonian}
\end{equation}
We have neglected the linear $k_i$ terms in the valence-valence
interaction, the quadratic $k_ik_j$ and $\varepsilon_{ij}$ terms in the
conduction-valence interaction and the ${k_i}{\varepsilon_{jk}}$ in the
strain-induced interactions. To simplify the analysis, we have imposed
axial symmetry in all the terms involved in the Hamiltonian. This 
approximation also affects the strain distribution forcing us to 
consider the materials as elastically isotropic. Compact expressions
are obtained with the Eshelby's inclusions method~\cite{eshelby_determination_1957}
and Fourier transform of the strain tensor~\cite{andreev_strain_1999}. 
With this formalism we succeeded in explaining Raman frequency shifts
induced by the strain~\cite{cros_raman_2006, garro_resonant_2006}
and strain distributions extracted from  middle energy ion scattering~\cite{jalabert_deformation_2005} experiments.
The solution of the problem is obtained considering that only the
band edges are discontinuous across material interfaces. To describe
electrons and holes properly in type-II QDs, we have decoupled the 
conduction and valence bands in $H$. Hence, the electrons are
described by a single band model, while a $6\times6$ Hamiltonian is used
for the holes. 

Axial symmetry allows us to define a total angular momentum of $z$
component $M$, which is the result of adding the corresponding
components of the envelope function orbital angular momentum ($m$) and
Bloch's amplitude total angular momentum ($j_z$): $M=m + j_z$. 
Each electronic state is described by the wave function
\begin{equation}\label{EQ:Psi}
\begin{split}
        \Psi^{(M)}(\vec{r}) & = \sum_{k=1}^8 F_{m, k}^{(M)}(\vec{r}) u_k \\
        &= \sum_{k=1}^8 \frac{1}{\sqrt{2\pi}} e^{i m \phi} \mathcal{F}_{m,k}^{(M)}(\rho,z)
        u_k,
\end{split}
\end{equation}
where $u_k$ is the Bloch amplitude of the $k$ band at the origin of 
the Brillouin zone and $F_{m,k}^{(M)}(\vec{r})$
is the envelope function associated to the $k$ Bloch component. 
The envelope function in expressed cylindrical coordinates defined by a 
phase factor characterized by $m$ and a two dimensional component 
$\mathcal{F}_{m,k}^{(M)}(\rho,z)$.

The solution of the Schr\"odinger equation is obtained by expanding
Eq.~\eqref{EQ:Psi} in a complete basis. The basis is defined by the
eigenfunctions of a hard-wall cylinder:
\begin{multline}
        \mathcal{F}^{(M)}_{m,k} = \sum_{\alpha,\mu} \mathcal{N}^{(m)}_\alpha
        J_m\left(k_\alpha^{(m)}\rho/\mathcal{R}\right)\\
        \times\sqrt{2/\mathcal{Z}}
        \sin\left[\mu\pi\left(z/\mathcal{Z} - 1/2\right)\right],
        \label{EQ:Expansion}
\end{multline}
where $J_m(x)$ is the Bessel function of order $m$, $k_\alpha^{(m)}$ is
its zero number $\alpha=1,2,\ldots$, $\mu=1,2,\ldots$, 
$\mathcal{R}$ and $\mathcal{Z}$ are the radius and height of the 
expansion cylinder and 
\begin{equation}
        N_\alpha^{(m)}=\frac{\sqrt{2}}{\mathcal{R}\left|J_{m+1}
        \left(k_\alpha^{(m)}\right)\right|}
        \label{EQ:Nomarlization}
\end{equation}
is the normalization of the radial part. This definitions ensure the
orthonormality of the expansion basis. A similar procedure was followed by
Tadić \emph{et al.} in Refs.~\onlinecite{tadic_effect_2002,tadic_intersublevel_2005}. Further details can be found in Ref.~\onlinecite{LlorensThesis}.

\begin{figure}[t]
        \includegraphics[scale=0.65]{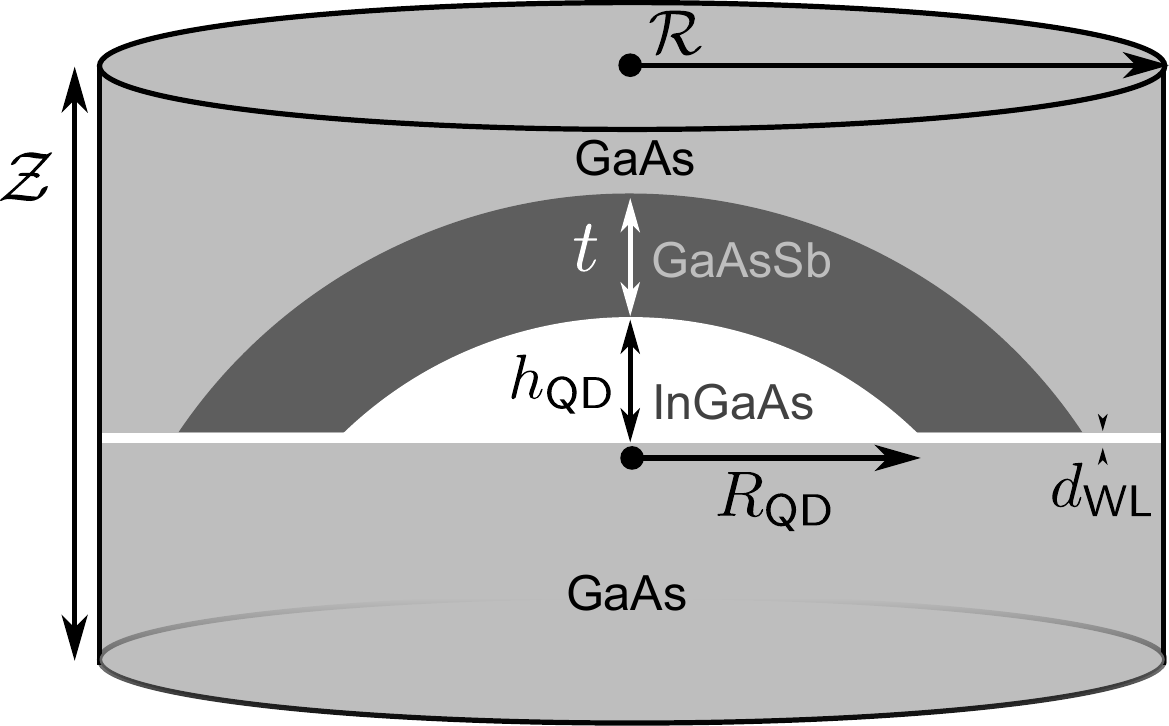}
        \caption{Depiction of the geometrical model to describe a QD of
                radius $R_\text{QD}$ and height $h_\text{QD}$, a overlayer of thickness
                $t$ and a wetting-layer of thickness $d_\text{WL}$
                embedded in hard-wall cylinder of radius $\mathcal{R}$
        and height $\mathcal{Z}$.}
        \label{fig:overlayer}
\end{figure}

In Figure~\ref{fig:overlayer}, we show an outline of the QD embedded in the
overlayer enclosed by the expansion cylinder. The number of geometrical parameters is therefore reduced to six. 

\section{Details of the EMA non-axisymmetric model}
\label{App:EMA}

The solution for the 3D Schr\"{o}dinger equation,
$\Phi(\rho,\theta,z)$, corresponding to the potential profile in Eq.~\eqref{inplanepotential} is given by
\begin{equation}\label{sol0}
\psi_{n,m,l}^{(e/h)}(\rho,\varphi,z)=\phi_{n,m}^{(e/h)}(\rho,\varphi)\chi_{l}^{(e/h)}(z)u_{e/h},
\end{equation}
where $\chi_{l}^{(e/h)}(z)$ is the wave function for a rigid square well and $u_{e/h}=|j,m_{j} \rangle$ are the
basis functions at the Brillouin zone center in the Kane model: $|1/2,\pm1/2\rangle$ and $| 3/2,\pm 3/2 \rangle$
for the electron and heavy hole, respectively. The
planar wave function has the form
\begin{align}
\phi_{n,m}^{(e/h)}(\rho,\varphi)&=\frac{1}{\lambda_{(e/h)}}\left(\frac{\Gamma[n+\mathcal{M}_{(e/h)}+1]}{2^{\mathcal{M}_{(e/h)}}n!(\Gamma[\mathcal{M}_{(e/h)}+1])^{2}}\right)^{1/2}\\\nonumber
&\times\left(\frac{\rho}{\lambda_{(e/h)}}\right)^{\mathcal{M}_{(e/h)}}\frac{e^{-im\varphi}}{\sqrt{2\pi}}e^{-\frac{1}{4}\left(\frac{\rho}{\lambda_{(e/h)}}\right)^2}
\\\nonumber
&\times
_{1}F_{1}\left(-n,\mathcal{M}_{(e/h)}+1,\frac{1}{2}\left(\rho/\lambda_{(e/h)}\right)^2\right),
\end{align}
where $_{1}F_{1}$  is the confluent hypergeometric function, $n=0,1,2,\ldots$ is
the radial quantum number, $l=1,2,\ldots$ is the vertical quantum number, and $m=0,\pm1,\pm2,\ldots$ labels the angular
momentum. The corresponding eigen-energies for the 3D problem are
\begin{multline}
E_{n,m,l,s_z}^{(e/h)} =\left(n+\frac{1}{2}+\frac{\mathcal{M}_{(e/h)}}{2}\right)\hbar{\omega_{(e/h)}} \\
-\frac{m}{2}\hbar{\omega_{c(e/h)}^*} -\frac{\mu_{(e/h)}^*}{4}{\omega_{0(e/h)}}^{2}{R_\text{QR}}^2 \\
+\left(\frac{l^{2}\pi^{2}\hbar^2}{2\mu_{(e/h)}^*{L^2}}\right) + g^*_{(e/h)} \mu_B B\cdot s_z,
\end{multline}
with $\mathcal{M}_{(e/h)}=\sqrt{m^2+\frac{2a_{1}\mu_{(e/h)}^*}{\hbar^2}}$,
$\omega_{c(e/h)}^*=eB/\mu_{(e/h)}^*$,
$\omega_{0(e/h)}=\sqrt{8a_{2}/\mu_{(e/h)}^*}$,
$\omega_{(e/h)}=\sqrt{\omega^{2}_{c(e/h)}+\omega^{2}_{0(e/h)}}$ and
$\lambda_{(e/h)}=\sqrt{\frac{\hbar}{\mu_{(e/h)}^*\omega_{(e/h)}}}$.

\section{Details of the SOC non-axisymmetric model}
\label{App:SOC}
The expression for the spin orbit Hamiltonian assuming the confinement profile that includes all the asymmetry terms is given by:
\begin{widetext}
        \begin{equation}\label{so}
                \begin{split}
                        H_{R}^c = 2\alpha_{c}\sigma_z\left[\left(-\frac{a_1}{\rho^2}+a_{2}\rho^2+\delta\rho^2\cos^2(\varphi)\right)\left(\frac{eB}{2c\hbar}-\frac{i}{\rho^2}\frac{\partial}{\partial\varphi}\right)
        -i\delta\sin(\varphi)\cos(\varphi) \left(1 + \rho \frac{\partial}{\partial\rho}\right)\right]\\
-\alpha_{c}\frac{\partial{V}}{\partial{z}}\left\{\sigma_{+}\left[e^{-i\varphi}\left(\frac{\partial}{\partial\rho}-\frac{i}{\rho}\frac{\partial}{\partial\varphi}+\frac{eB}{2c\hbar}\rho+\frac{1}{\rho}\right)\right]\right.
\left.-\sigma_{-}\left[e^{i\varphi}\left(\frac{\partial}{\partial\rho}+\frac{i}{\rho}\frac{\partial}{\partial\varphi}-\frac{eB}{2c\hbar}\rho+\frac{1}{\rho}\right)\right]\right\},
                \end{split}
        \end{equation}
\end{widetext}
with $\sigma_{\pm}=1/2(\sigma_x \pm \sigma_y)$, being $\sigma_i$ the components of the Pauli matrices. The spin orbit Hamiltonian for the valence band can be readily obtained by replacing these matrices with those for angular momentum $j=3/2$ and $\alpha_{c}$ by $\alpha_{v}$. 

We can extract from Eq.~\eqref{so} the diagonal terms that contribute to the
first order renormalization of the spin-splitting of the ground state with
$m=0$ for the limit $B\rightarrow 0$, $\Delta E_\text{spin}=\Delta E^{(0)} + \Delta E^{(1)}$,
in the basis of unperturbed states introduced by Eq.~\eqref{sol0}. The zero
order value is essentially the Zeeman splitting given by
\begin{equation}\label{de10}
\Delta E_c^{(0)}=g^* \mu_B B,
\end{equation}
in the case of the conduction band, while for the valence band stands as
\begin{equation}\label{dev0}
\Delta E_v^{(0)}=-6 \kappa \mu_B B.
\end{equation}
In the case of a conduction band electron confined within a QD potential, $V=a_2 \rho^2+\delta \cdot \rho^2 \cos^2{\varphi}$, the first order contribution of the spin-orbit interaction, in the limit of low fields, can be reduced to
\begin{equation}\label{de1}
        \begin{split}
\Delta E_c^{(1)}&=\frac{4 m_0 \alpha_{c}}{\hbar^2}\\
&\times  a_2 \left\{ 1+ \text{sign}(\delta) \frac{e^2}{2- \left[1+ \text{sign}(\delta)\right]e^2}   \right\} \left\langle \rho^2  \right\rangle  \mu_B B.
        \end{split}
\end{equation}
Meanwhile, for the valence band and a quantum ring profile, $V=a_1/\rho^2
+a_{2}\rho^2-2\sqrt{a_{1}a_2}+\delta \cdot \rho^2 \cos^2{\varphi}$, the expression reads
\begin{equation}\label{dv1}
        \begin{split}
\Delta E_v^{(1)}&=\frac{6 m_0 \alpha_{v}}{\hbar^2}\\
&\times\left[  a_2 \left\{ 1+ \text{sign}(\delta) \frac{e^2}{2- \left[1+ \text{sign}(\delta)\right]e^2}   \right\} \left\langle \rho^2  \right\rangle  \right.\\
&\left.- a_1 \left\langle \frac{1}{\rho^2}  \right\rangle \right]\mu_B B.
        \end{split}
\end{equation}

This allows introducing the
first order correction to the Land\'{e} factor defined as $g_{c,v}^{(1)} \equiv \Delta
E_{c,v}^{(1)}/ (\mu_B B)$ and shown in Fig. \ref{figSOBC} and \ref{figSOBV} in the article, respectively. With them, one may now assess the relative effect of the
confinement geometry and topology change on the spin splitting. 
The calculation of the Rashba coefficient appearing in Eq.\eqref{dv1} is computed from the band parameters in Table \ref{tab:magnetic_params} and the expression:
\begin{equation}
\alpha_v = -\dfrac{eP^2}{3E_0^2}+\dfrac{eQ^2}{9}
\left[ \dfrac{10}{{E'}_0^2}-\dfrac{7}{(E_0'+\Delta_0')^2}\right],
\end{equation}
which is defined as $r_{41}^{8v8v}$ in \onlinecite{winkler_book}.

\section{Material parameters}
\label{App:Mat}
The material band parameters are extracted from \cite{vurgaftman2001}. Magnetic
related parameters are shown in Table~\ref{tab:magnetic_params}. The values of the compounds GaInAs and GaAsSb are obtained through linear interpolation when no bowing parameters is
reported.

\begin{table}[b]
        \begin{tabular}{r r r r}
                & GaAs & InAs & GaSb \\\hline\hline
                $g$ & -0.44 & -14.9 & -9.25 \\
                $k$ & 1.20 & 7.60 & 4.60 \\
                $q$ & 0.01 & 0.39 & 0.00 \\
          $E_0$ (eV) & 1.519 & 0.418 & 0.813 \\
          $E_0'$ (eV) & 4.488 & 4.390 & 3.3 \\
          $\Delta_0$ (eV) & 0.341  & 0.380 & 0.750 \\
          $\Delta_0'$ (eV) & 0.171  & 0.240 & 0.330 \\
          $P$ (eV\AA) & 10.493 & 9.197 & 9.504 \\
          $P'$ (eV\AA) & 4.780$i$ & 0.873$i$ & 3.326$i$ \\
          $Q$ (eV\AA) & 8.165 & 8.331 & 8.121  \\\hline\hline
        \end{tabular}
        \caption{Parameters related with the spin Zeeman effect and band parameters
        used in the computation of the Rashba coefficient. The values
                of GaAs and InAs are taken from Ref.~\onlinecite{winkler_book} (page 221). The
                Zeeman parameters of GaSb from Ref.~\onlinecite{Meier1984} (pages 486 and 491)
                and the band parameters from Ref.~\onlinecite{cardona_relativistic_1988}. }
        \label{tab:magnetic_params}
\end{table}

%

\end{document}



\title{Supplementary Material of topology driven g-factor tuning in type-II quantum dots}


\author{J. M. Llorens}
\affiliation{Instituto de Micro y Nanotecnología, IMN-CNM, CSIC (CEI UAM+CSIC) Isaac Newton, 8, E-28760, Tres Cantos, Madrid, Spain}

\author{V. Lopes-Oliveira}
\affiliation{Universidade Estadual de Mato Grosso do Sul, Dourados, Mato Grosso do Sul 79804-970, Brazil}

\author{V. López-Richard}
\affiliation{Departamento de Física, Universidade Federal de São Carlos, São Carlos, São Paulo 13565-905, Brazil}

\author{E. R. Cardozo de Oliveira}
\affiliation{Departamento de Física, Universidade Federal de São Carlos, São Carlos, São Paulo 13565-905, Brazil}

\author{L. Wewiór}
\affiliation{Instituto de Micro y Nanotecnología, IMN-CNM, CSIC (CEI UAM+CSIC) Isaac Newton, 8, E-28760, Tres Cantos, Madrid, Spain}

\author{J. M. Ulloa}
\affiliation{Institute for Systems based on Optoelectronics and Microtechnology (ISOM), Universidad Politécnica de Madrid, Ciudad Universitaria s/n, 28040 Madrid, Spain}

\author{M. D. Teodoro}
\affiliation{Departamento de Física, Universidade Federal de São Carlos, São Carlos, São Paulo 13565-905, Brazil}

\author{G. E. Marques}
\affiliation{Departamento de Física, Universidade Federal de São Carlos, São Carlos, São Paulo 13565-905, Brazil}

\author{A. García-Cristóbal}
\affiliation{Instituto de Ciencia de Materiales (ICMUV), Universidad de Valencia, Paterna E-46980, Spain}

\author{G.-Q. Hai}
\affiliation{Instituto de Física de São Carlos, Universidade de São Paulo, São Carlos, São Paulo 13560-970, Brazil}

\author{B. Alén}
\affiliation{Instituto de Micro y Nanotecnología, IMN-CNM, CSIC (CEI UAM+CSIC) Isaac Newton, 8, E-28760, Tres Cantos, Madrid, Spain}
\email[E-mail address:]{benito.alen@csic.es}


\date{\today}

\pacs{}
\begin{abstract}
\end{abstract}

\maketitle

\renewcommand{\thefigure}{S\arabic{figure}}
\setcounter{figure}{0}
\renewcommand{\thetable}{S\arabic{table}}
\setcounter{table}{0}

\section{MPL spectral analysis details}
\begin{figure*}[h]
\includegraphics[scale=.85]{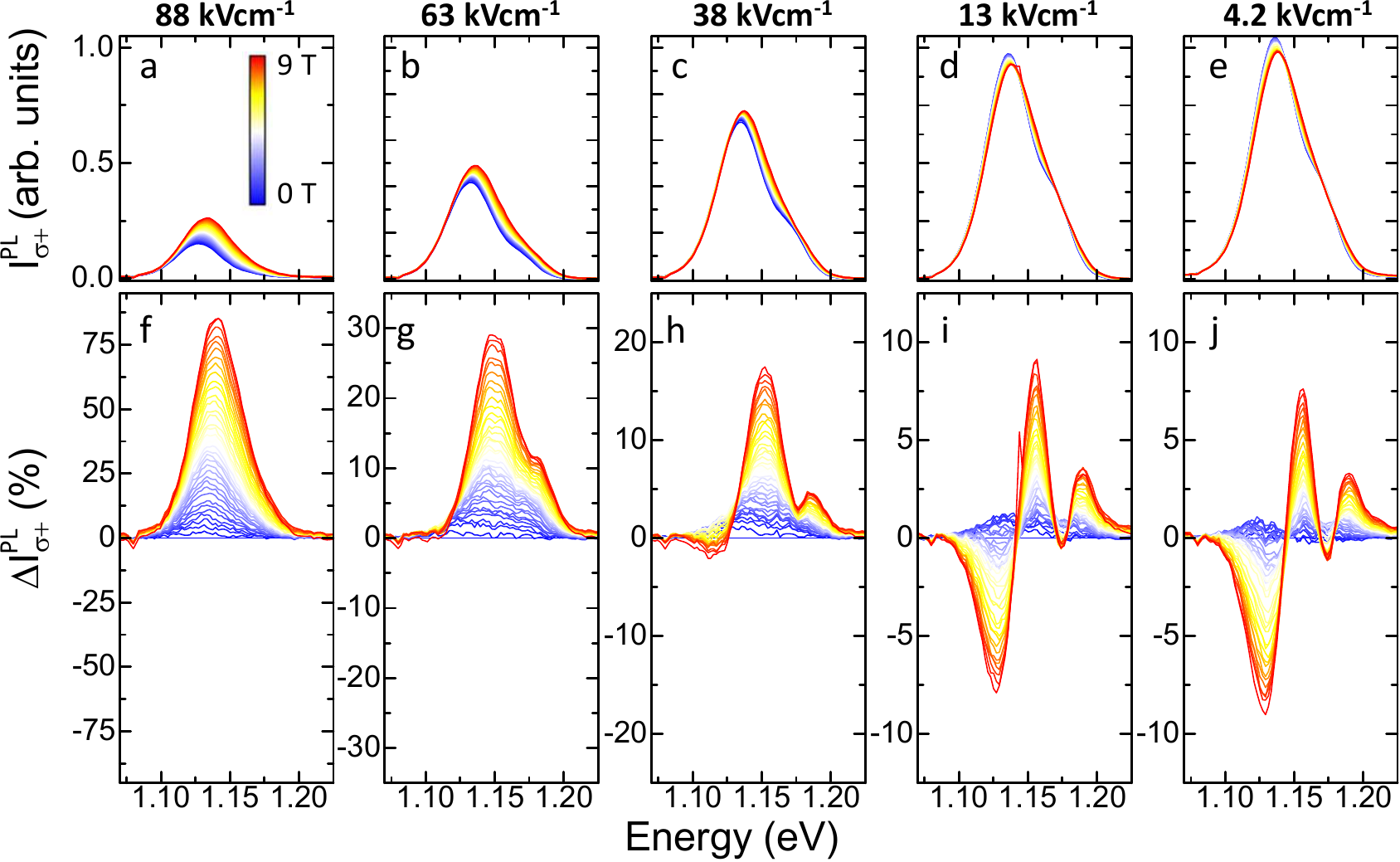}
%
%
\caption{(a-e) Evolution of the $\sigma^+$-polarized magneto-photoluminescence in vertical electrical fields between 88 and 4.2 kV cm$^{-1}$. To highlight the magnetic field effects, (f-j) show the evolution after subtraction and normalization by the spectrum at 0 T.}
\label{fig:MPL_smas}       
\end{figure*}

\begin{figure*}[h]
\includegraphics[scale=.65]{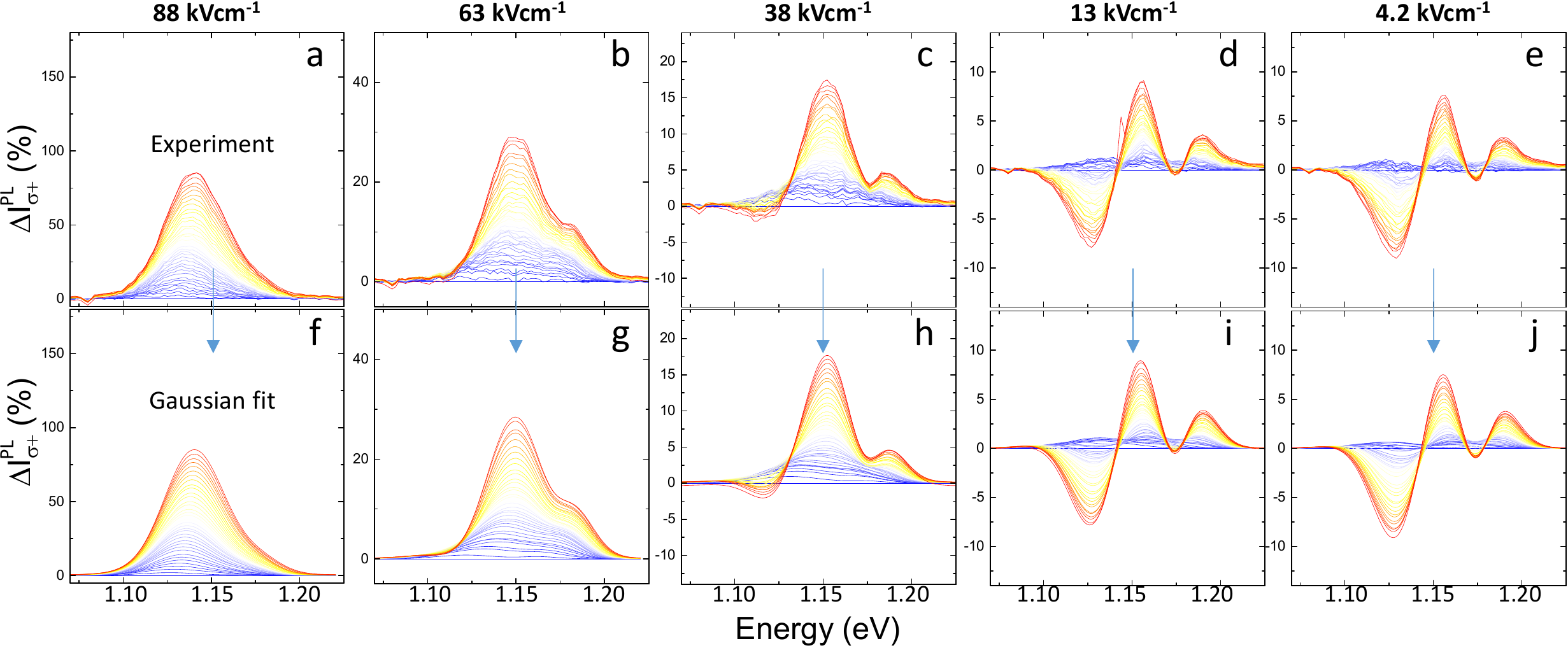}
%
%
\caption{(a-e) Evolution of the $\sigma^+$-polarized magneto-photoluminescence in vertical electrical fields between 88 and 4.2 kV cm$^{-1}$ after subtraction and normalization by the spectrum at 0 T. (f-j) Same representation for the spectra obtained after Gaussian deconvolution of the data.}
\label{fig:MPL_comp}       
\end{figure*}

\begin{figure*}[h]
\includegraphics[width=0.90\textwidth]{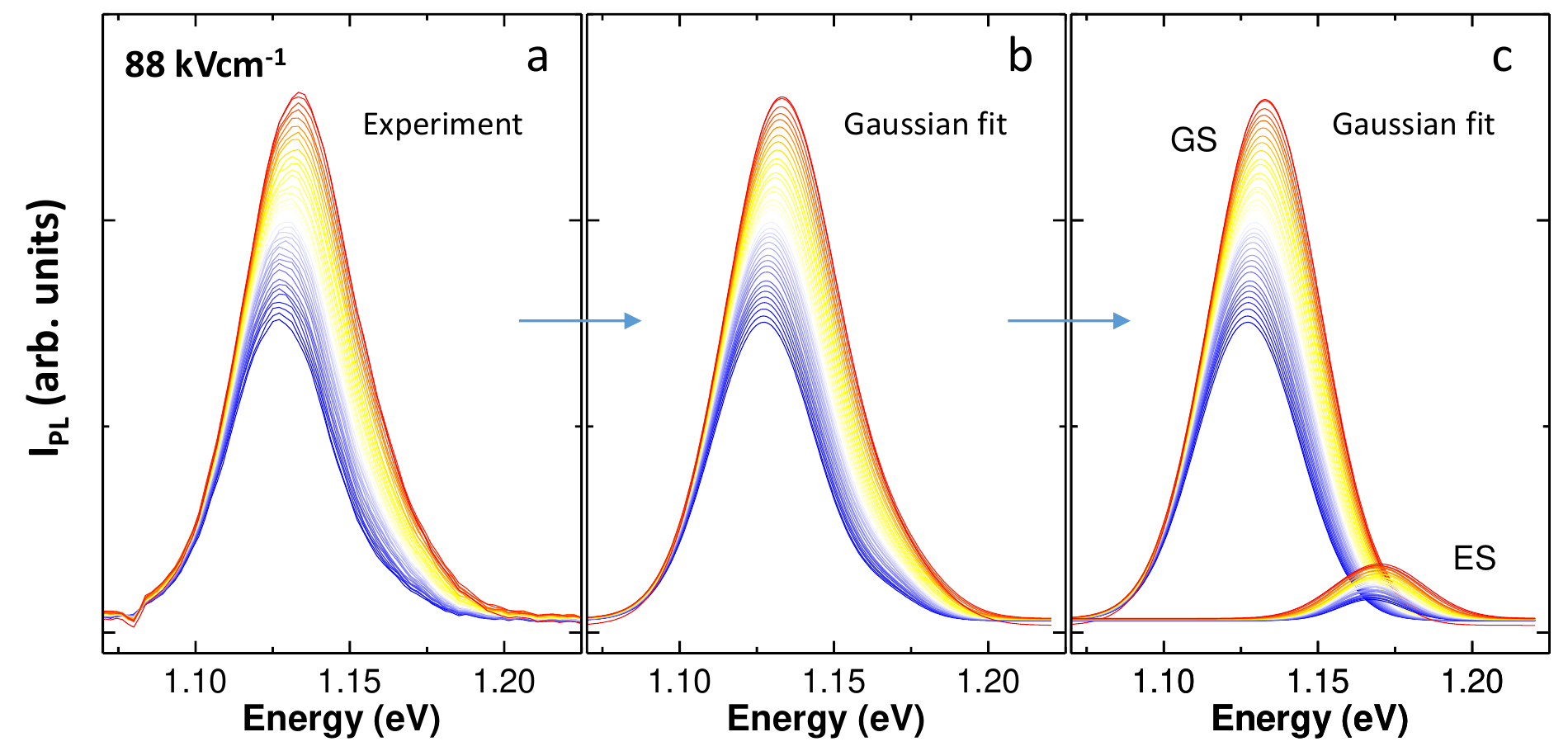}
%
%
\caption{(a) Evolution of the $\sigma^+$-polarized magneto-photoluminescence at 88 kV cm$^{-1}$. (b,c) Results of the two-component Gaussian deconvolution procedure.}
\label{fig:MPL_decon}       
\end{figure*}

Figure~\ref{fig:MPL_smas} shows the $\sigma^+$-polarized MPL spectra which completes the $\sigma^-$ data shown in Figure~2 of the manuscript. Every spectrum was spectrally analyzed by two-component Gausssian deconvolution. To improve the fidelity of the analysis, the fit to the experimental data was done after subtraction and normalization by the spectrum at 0 T, as shown in figure ~\ref{fig:MPL_comp}. Figure~\ref{fig:MPL_decon} shows the result of deconvolution with ground state (GS) and excited state (ES) explicitly indicated. The low energy band (GS) was then further analyzed in the discussion of the magnetic properties of the InAs/GaAsSb system.

%

%

%


\section{Intensity emission calculation with a rate equations model}

\begin{figure}[h]
\includegraphics[scale=.65]{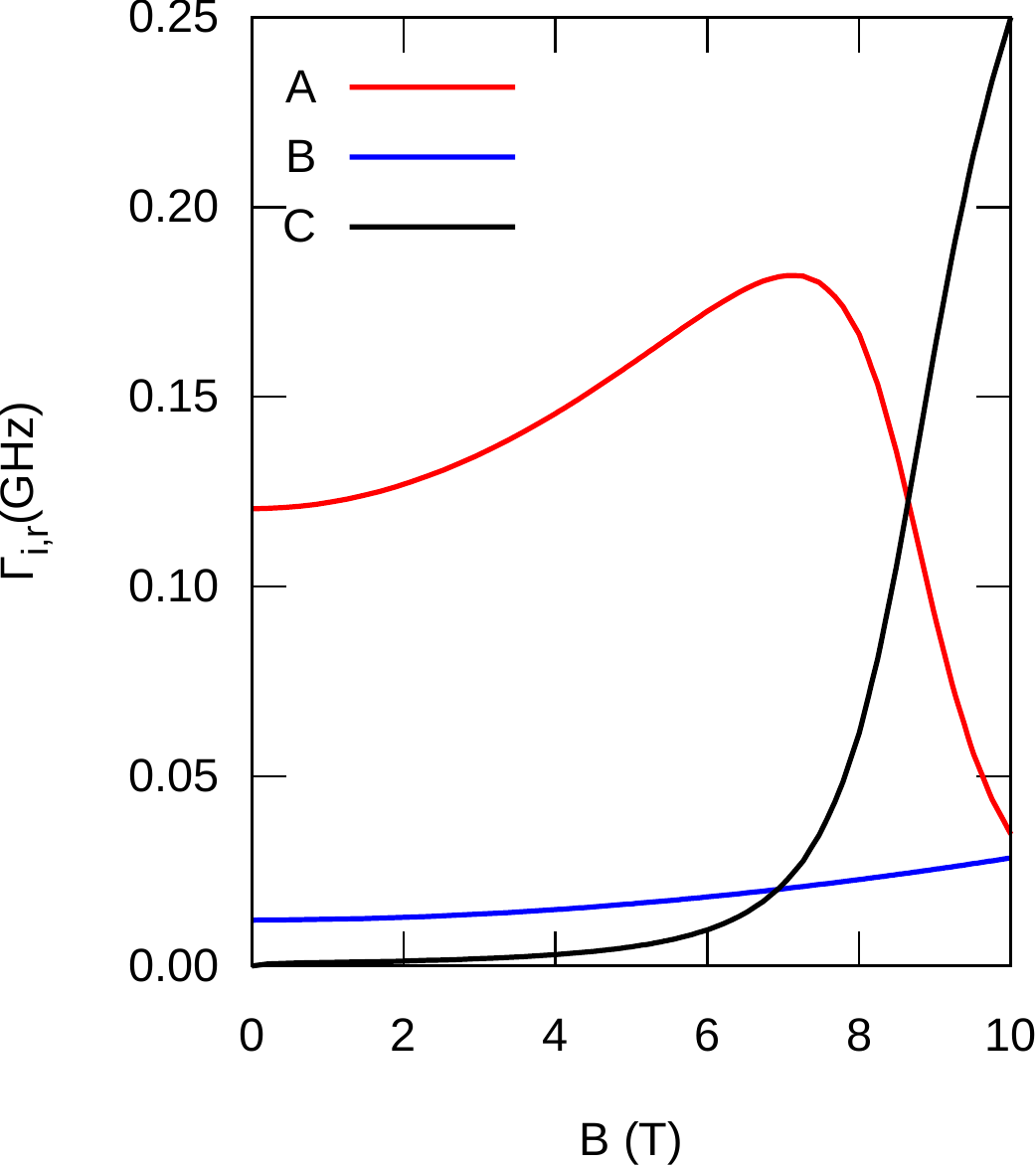}
%
%
\caption{Radiative rate of states A, B and C considered in the calculation.}
\label{fig:rates}       
\end{figure}

For the computation of the emission intensity, we simplify the electronic structure around the QD ground state as being described by the three levels ``A'', ``B'' and ``C'' of Fig. 7 in the main text. The radiative rate of each state is proportional to the oscillator strength of Fig. 7(c). The rate equations model is based on Ref.~\onlinecite{sugawara_quantumdot_2002}. The first order differential equations system is:

\begin{align}
        \dfrac{dN_\text{R}}{dt} &= G - (\Gamma_{\text{R}\rightarrow \text{C}} + \Gamma_{\text{R},r}) N_\text{R}\\
\dfrac{dN_\text{C}}{dt} &= \Gamma_{\text{R}\rightarrow \text{C}} N_\text{R} + \Gamma_{\text{B}\rightarrow \text{C}} N_\text{B} - (\Gamma_{\text{C},r}  + \Gamma_{\text{C}\rightarrow \text{B}}) N_\text{C}\\
        \dfrac{dN_\text{B}}{dt} &= \Gamma_{\text{C}\rightarrow \text{B}} N_\text{C} +\Gamma_{\text{A}\rightarrow \text{B}} N_\text{A} - (\Gamma_{\text{B},r} + \Gamma_{\text{B}\rightarrow \text{A}}  + \Gamma_{\text{B}\rightarrow \text{C}}) N_\text{B} \\
        \dfrac{dN_\text{A}}{dt} &= \Gamma_{\text{B}\rightarrow \text{A}} N_\text{B} - (\Gamma_{\text{A},r} + \Gamma_{\text{A}\rightarrow \text{B}}) N_\text{A}
\label{EQ:rate}
\end{align}
where $N_i$ is the population of a state $i$. The transition rate to a state of lower energy is proportional to the occupation of that state $\Gamma_{i\rightarrow j} = (1 - P_j)\Gamma_{i\rightarrow j}^0$. The occupation of a state $j$ is the ratio between its population and the degeneracy $P_j = N_j/g_j$. The degeneracy is two for the three states.  The thermal excitation of carriers from lower towards higher energy levels $\Gamma_{j\rightarrow i} = (1 - P_i) \Gamma_{j\rightarrow i}^0$. Under the aproximation of the detailed balance between relaxation and emission rate: $\Gamma_{j\rightarrow i}^0 = \Gamma_{i\rightarrow j}^0 \exp[-(E_j - E_i)/k_\text{B}T]$ where $k_\text{B}$ is the Boltzmann constant and $T$ the temperature. Since the ensemble experiment cannot resolve these individual states, the emitted intensity is computed as the sum of the radiative rate times the population of the three states. We take as parameters vaues $\Gamma_{i\rightarrow j}^0$ = $\Gamma_{\text{R}\rightarrow\text{C}}$ = 100 GHz, $T$ = 5 K, $\Gamma_{\text{R},r}$ = 1 GHz. The maximum radiative rate of the states is $\Gamma_{r,\text{max}}$ = 0.25 GHz. The values of $\Gamma_{\text{A},r}$ and $\Gamma_{\text{C},r}$ are scaled according to the computed wave-function overlap in Fig.~7(c).
In the case of state B, pure dark conditions lead to unphysical charge accumulation effects within the recombination model. We relax this condition assuming that state B can recombine radiatively with a lifetime $\Gamma_{\text{B},r}$ = $(\Gamma_{\text{A},r}+\Gamma_{\text{C},r})/10$, i.e. an order of magnitude smaller that the rates of A and C. This radiative coupling shall arise from hybridization with bright states beyond the current electronic structure model. Under these approximation, the calculated emission intensity (represented in Fig.~8) oscillates around the AB field as experimentally observed. The resulting rates are shown in Fig.~\ref{fig:rates}. 

\bibliography{sup_mat}